\documentclass[journal=jacsat,manuscript=article, layout=onecolumn]{achemso}

\usepackage{chemformula} 
\usepackage[T1]{fontenc} 

\usepackage{color}
\usepackage{xcolor}
\usepackage{float}
\usepackage{mathpazo}
\usepackage{hyperref}
\usepackage{xr}

\definecolor{scarlet}{HTML}{BA0C2F}
\newcommand{\vect}[1]{\boldsymbol{#1}}
\author{Diego Becerra}
 \email{becerra.39@osu.edu}
 \affiliation{William G. Lowrie Department of Chemical and Biomolecular Engineering, 
The Ohio State University, Columbus, Ohio 43210, USA}

\author{Pranav R. Jois}
 \affiliation{Department of Mathematics and Department of Physics, The Ohio State University, Columbus, Ohio 43210,
USA
}

\author{Lisa M. Hall}
 \email{hall.1004@osu.edu}
 \affiliation{%
William G. Lowrie Department of Chemical and Biomolecular Engineering, 
The Ohio State University, Columbus, Ohio 43210, USA
}%

\title[]{Coarse-grained modeling of polymers with end-on and side-on liquid crystal moieties: 
effect of architecture
}

\abbreviations{}

\begin{document}

\newpage
\begin{abstract}
 Mesogens, which are typically stiff rodlike or disklike molecules, are able to self-organize into liquid crystal (LC) phases in a certain temperature range. Such mesogens, or LC groups, can be attached to polymer chains in various configurations including within the backbone (main-chain LC polymers) or at the ends of side-chains attached to the backbone in an end-on or side-on configuration (side-chain LC polymers or SCLCPs), which can display synergistic properties arising from both their LC and polymeric character. At lower temperatures, chain conformations may be significantly altered due to the mesoscale LC ordering, thus, when heating from the LC ordered state through the LC to isotropic phase transition, the chains return from a more stretched to a more random coil conformation. This can cause macroscopic shape changes, which depend significantly on the type of LC attachment and other architectural properties of the polymer. Here, to study the structure-property relationships for SCLCPs with a range of different architectures, we develop a coarse-grained model that includes torsional potentials along with LC interactions of a Gay--Berne form. We create systems of different side chain lengths, chain stiffnesses, and LC attachment types, and track their structural properties as a function of temperature. Our modeled systems indeed form a variety of well-organized mesophase structures at low temperatures, and we predict higher LC to isotropic transition temperatures for the end-on side-chain systems than for analogous side-on side-chain systems. Understanding these phase transitions and their dependence on polymer architecture can be useful in designing materials with reversible and controllable deformations.
\end{abstract}

\newpage

\section{INTRODUCTION}
\label{sec:introduction}
Liquid crystals (LCs) are stiff self-organizing molecules 
that can be attached to relatively flexible polymer chains
either within the backbone or in 
an end-on or side-on configuration along the backbone, yielding a material that takes advantage of the properties of both types of materials.
The backbone tends towards a random walk at high temperatures, 
but when the LC groups order at lower temperatures, the
backbone is distorted and the material expands or contracts (depending on LC attachment) 
along the LC director. 
Though our focus here is on uncrosslinked systems, LC polymers can be lightly crosslinked to take greater advantage of such deformations on a macroscopic scale. Such LC elastomers (LCEs), with LC moieties
attached to the chains in various ways, have been the focus of significant recent research
\cite{allen1989computer,lu2017tunable,yang2018multitemperature}. 
Importantly, due to the stimuli-responsive nature of the LC groups, or mesogens, LCEs can be placed in a state with kinetically trapped polymer conformations 
with low conformational entropy, 
and they can later be prompted to return to their original state by an external stimulus 
\cite{bellin2006polymeric,ahir2006self,qin2009combined,ware2016localized,xu2021random}.
These types of shape changes can be leveraged in applications such as artificial muscles, as first proposed by de Gennes 
\cite{de1969possibilites}.
With synthetic advances that allow for additional complexity and tunability in their deformation behavior, 
LCEs can now allow for a variety of controlled and
reversible thermo- and photo-mechanical responses \cite{warner2007liquid,behl2007shape,ahn2011side}, 
of interest in 
soft robotics, optical data storage, biomedicine, 
and tissue engineering
\cite{shenoy2002carbon,spillmann2007stacking,
li2006artificial,kim2019contactless,wang2022multi,
kempe2004self,kempe2006rheological,wang2022multi}.

Here, we model simplified systems of LC polymers 
(that are not cross-linked to make LCEs), 
focusing on how polymer architectural parameters lead to different microstructures 
that would then lead to different responses if made into an LCE material.
In general, LCs can be attached to polymer chains
either within the backbone, leading to main-chain LC polymers (MCLCPs), 
or in an end-on or side-on configuration at the ends of short polymeric segments 
(also called ``spacers'') attached to the backbone, 
leading to side-chain LC polymers (SCLCPs).
These spacers usually have a small number of repeat units and
allow the polymer main backbone to accommodate 
the anisotropic orientation of mesogenic side groups \cite{thakur2016liquid}, 
allowing for coupling/decoupling the constituent parts.
Although these two types of LC polymers share conformational units, they are intrinsically different;
compared with the well-studied MCLCPs 
\cite{moore1987orientation,affouard1996molecular,
lyulin1998molecular,chen2016gpu,cuierrier2021simulation,berardi2004computer}, 
where mesogenic groups are the monomers that compose the polymer itself, 
in the SCLCPs, 
the structure of the polymer backbone, 
the grafting density of side chains, 
the LC type of attachment,
the tacticity,
and the spacer length 
all exert an important influence 
on the mesophase morphology 
\cite{stimson2005molecular,collings2017introduction}. 

At high enough temperatures,
thermotropic LC polymers offer typical polymeric features 
such as elasticity, and their
polymer chains approach a random coil configuration 
due to entropic considerations. 
However, at lower temperatures, 
the LC groups form ordered phases 
(e.g., nematic or smectic). 
This LC ordering occurs at the expense of lowering 
the polymeric conformational entropy, and the
conformations of the flexible, non-LC polymer segments 
can be significantly stretched. 

In this context, understanding the morphology, 
dynamics, and self-organization in SCLCP systems 
is of great fundamental and technological interest 
as has been demonstrated over the years through 
both experimental
\cite{kannan1993rheology,d1988experimental,
wewerka2001side,wewerka2001structure,rendon2007shear,colby1993linear}
and theoretical
\cite{pleiner1992local,fourmaux1998rheology,long2002rouse,
wang1987theory,renz1988theory,brochard1984phase,chiu1995equilibrium,
kim2007phase,carri1998configurations} 
studies, which have studied these systems' phase diagrams, 
elastic properties, atomistic structuring, 
hydrodynamics, and classical rheology. A main characteristic of SCLCP systems, and one which makes them challenging to model, is the 
existence of orientational and possibly positional order of their constituent parts along with molecular scale features on different length scales, 
as well as the presence of transitions and sometimes coexistence between different phases. 
In this regard, simplified models can be useful in understanding the overall phase behavior of SCLCPs. In particular, \citet{wang2010theory} applied a self-consistent-field theoretic approach to SCLCPs; their model takes into account the coupling
between the orientation of the side-chain LC groups and that of the backbone segments, 
both locally and globally.
One of the main takeaways is that for side-on SCLCPs, 
the coupling effects act cooperatively so that the chain conformation is always prolate, 
meaning that the polymer chains align in the same direction as the LC groups in the LC mesophases. 
Instead, for the end-on SCLCPs, these effects act competitively and 
the chain conformation can be either prolate or oblate (meaning that the
polymer chains are stretched perpendicular to the LC director in the LC mesophases).

Since adequate descriptions of SCLCP systems require consideration of
both inter- and intra-molecular interactions
and also system sizes large enough to effectively capture overall molecular ordering,
coarse-grained molecular simulations are appropriate\cite{pasini2000liquid,wilson2022molecular}.
Some attempts to model SCLCPs 
in prior works are based on 
hybrid models of spherically symmetrical and anisotropic sites 
\cite{wilson1997molecular,stimson2005molecular,
mcbride1999molecular,wilson2003computer, 
ilnytskyi2007structure,ilnytskyi2012modelling,chen2016gpu,
cuierrier2021simulation}.
In particular \citet{stimson2005molecular} proposes a coarse-grained model 
for SCLCP molecules which arises as an extension 
of a series of previous works on liquid crystal polymers 
from the same group
\cite{wilson1997molecular,mcbride1999molecular,
wilson2003computer}. 
The model, which is meant to represent a specific chemical type of polymer with a 
polydimethylsiloxane backbone and flexible side chains terminated in mesogenic groups, 
includes harmonic bonds, dihedral potentials, 
Lennard-Jones (LJ) interactions for backbone beads, and anisotropic 
interactions for mesogenic groups (specifically, a Gay--Berne model 
\cite{gay1981modification,cleaver1996extension,allen1996molecular}). 
By applying a small aligning potential during a cooling processes 
from fully isotropic polymer melt, 
they found microphase formation, 
specifically, to a smectic-A phase, which is in good agreement 
with experimental scattering data.
Later simulations by Ilnytskyi $et\,al.$ provided further understanding 
of SCLCP systems
\cite{ilnytskyi2007structure,ilnytskyi2011opposite,ilnytskyi2012modelling}.
Using a similar modeling technique (semiatomistic level of description) 
as the previously mentioned group, they studied end-on SCLCP systems of
a range of backbone and 
spacer lengths, giving rise to 
isotropic, polydomain smectic, and monodomain smectic phases, 
the latter by using an external alignment field.  
Also, they performed structural and dynamic analyses  
from equilibrium and non-equilibrium simulations using their 
in-house MD code
\cite{ilnytskyi2001domain,ilnytskyi2002domain}.

In this work we also employ a coarse-grained model 
for SCLCPs, but with additional orientational interactions discussed below. 
We simulate both end-on SCLCP systems (as studied in previous simulation works) 
and side-on SCLCP systems (we believe for the first time) and analyze a range of structural phenomena.
We aim to understand the relationship 
between the molecular architecture, including LC type of attachment, 
and the morphology of any ordered phases.
Our underlying polymeric model 
is inspired by an elastic rod model 
for semiflexible polymers that accounts for bending, 
orientational, and twisting interactions 
\cite{chirico1994kinetics,brackley2014models,lequieu20191cpn}.
By including these orientations, 
making a fully anisotropic model even for the spherical polymeric beads, we can capture 
a broader range of possible molecular architectures,
including reproducing tacticity. 
Though the tracking of orientations and torques adds computational expense, 
 we can still simulate relatively large systems on long timescales.

The paper is organized as follows. 
We first discuss the details of the SCLCP coarse-grained model and 
the simulation methodology (Sec.~\ref{sec:model-methodology}). 
In Sec.~\ref{sec:results}, the
results obtained for 
different architectural parameters of the model and 
the corresponding mesophases formed in different thermodynamic states 
are presented and discussed. 
Finally, in Sec.~\ref{sec:performance} we show 
some comparative computational performance results.
The coarse-grained model is implemented in
the open-source LAMMPS package \cite{LAMMPS} 
and the particular external orientational alignment field discussed below
is freely available online \cite{becerra2022fixext}.

\section{MODEL AND METHODOLOGY}
\label{sec:model-methodology}

\subsection{Coarse-grained model}
\label{subsec:model}

In our coarse-grained model for SCLCPs, 
the polymer backbone is represented by a 
twistable worm-like chain model 
\cite{chirico1994kinetics,brackley2014models}, 
composed of spherical beads, 
to which pendant side chains are attached. In the current work, side chains are attached
syndiotactically, with a grafting density of one side chain 
every other backbone bead.
The side chains are composed of a number of spacer beads with the same model as backbone beads
followed by a terminal mesogenic group 
modeled as an ellipsoid using a Gay--Berne (GB) potential
\cite{gay1981modification,cleaver1996extension,allen1996molecular}, a generalization of the 12-6 Lennard-Jones (LJ) potential
for anisotropic interactions.
These LC moieties are attached in an end-on (with the long axis parallel to the last spacer bond)
or side-on (with the long axis perpendicular to the last spacer bond) configuration. 

\begin{figure}
	\centering
	\includegraphics[width=1\textwidth]{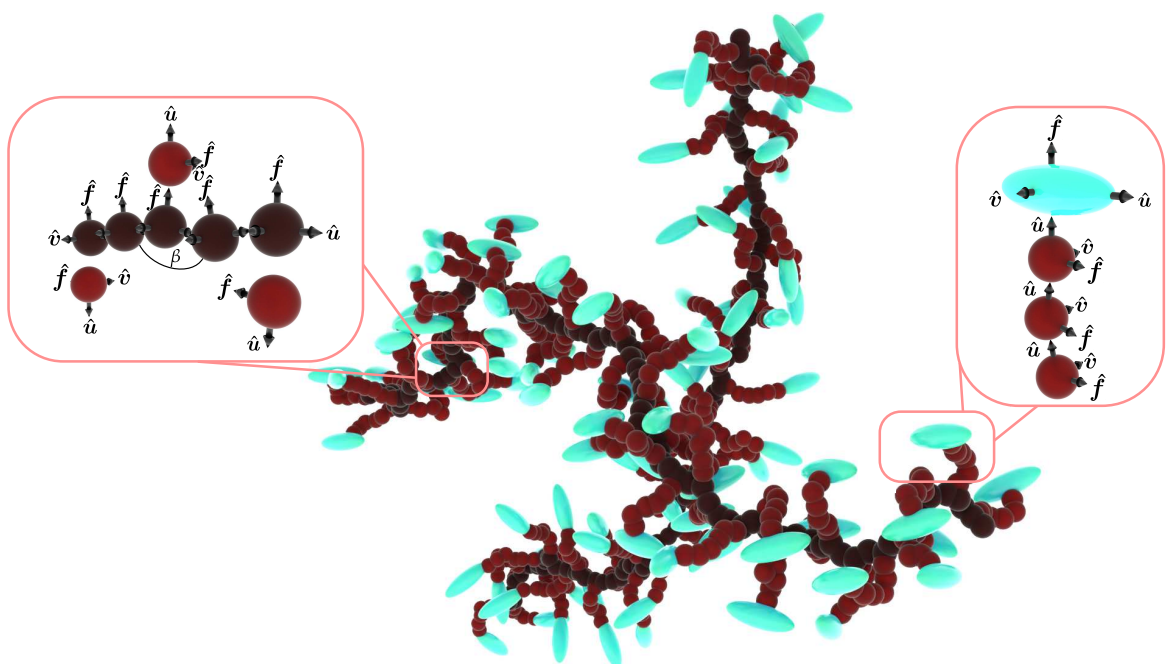}
	\caption{Representation of two side-chain liquid crystal polymer 
	molecules,
	one containing purely end-on mesogenic groups and the other 
	containing purely side-on mesogenic groups. 
	Liquid crystals are modeled by ellipsoidal sites (cyan) 
	attached to the spacer chains (scarlet), which in turn are attached to the polymer main backbone (dark scarlet).
	The left magnified image shows the first spacer beads 
	orientationally attached to the backbone whereas
	the right magnified image shows a side-on liquid crystal moiety attached
	to the respective spacer.
	The $\hat{\vect{f}},\,\hat{\vect{v}},\,\hat{\vect{u}}$ unit vectors 
	(arrows)
	denote the orientation of each anisotropic site.
	Images were rendered using Blender \cite{blender}.
	}
	\label{fig:sketch}
\end{figure}

Orientations and angular momenta are tracked for all sites/beads using quaternions for computational efficiency. 
However, for an intuitive representation of 
the model and its potentials, 
here we discuss the orientation of each site in 
terms of an typical orthonormal basis in 3D
\cite{brackley2014models,lequieu20191cpn}, with
the orientation of the $i$th site in the lab-frame being 
$(\hat{\vect{f}}_{i},\,\hat{\vect{v}}_{i},\,\hat{\vect{u}}_{i})$.
A representation of two SCLCP molecules with different
LC types of attachments is shown in
Figure~\ref{fig:sketch}. 

\subsubsection{Nonbonded interactions}

Pairwise interaction potentials exist between all nonbonded pairs of sites; since we use anisotropic sites, 
in some cases the interaction between particles $i$ and $j$
depends on both the distance and the orientation of particle 
$i$ or $j$ (or both). 
The specific interactions are:
(i) a typical 12-6 Lennard-Jones (LJ) potential, $U_{\text{LJ}}({\vect{r}}_{ij})$,
if both sites involved are spherical, 
i.e., $i$ and $j$ correspond to either backbone beads or 
spacer beads;
(ii) a Gay--Berne (GB) anisotropic potential \cite{gay1981modification}, 
$U_{\text{GB}}(\hat{\vect{u}}_{i},\,\hat{\vect{u}}_{j},\,{\vect{r}}_{ij})$,
if both sites involved in the interaction 
are aspherical, i.e., $i$ and $j$ correspond to LC moieties 
(we choose parameters to give prolate ellipsoids, longer in one axis than the others); and 
(iii) a cross term 
$U_{\text{GB/LJ}}(\hat{\vect{u}}_{i},\,{\vect{r}}_{ij})$,
which represents a generalized potential for two unlike Gay--Berne particles, 
if $j$ corresponds to a spherical site 
and $i$ to an aspherical site, 
or $U_{\text{GB/LJ}}(\hat{\vect{u}}_{j},\,{\vect{r}}_{ij})$, 
if $i$ corresponds to a spherical site 
and $j$ to an aspherical site.
Here, ${\vect{r}}_{ij}={\vect{r}}_{i}-{\vect{r}}_{j}$ is the 
interparticle vector, and
$\hat{\vect{u}}_{i}$ and $\hat{\vect{u}}_{j}$ are the unit vectors 
of the main axis of the mesogenic groups of sites $i$ and $j$, respectively.
Moreover, in this work, the parameters are described 
in dimensionless units in terms 
of the characteristic LJ length $\sigma$, energy $\varepsilon$, 
and mass $m$.
For all spherical interactions, 
the mass, LJ energy, and LJ diameter parameters are set to unity to simulate a generic polymer.
GB interactions \cite{gay1981modification,berardi1998gay} between two anisotropic particles or between 
pairs of dissimilar particles are calculated
using the formalism of \citet{everaers2003interaction} 
as implemented in LAMMPS \cite{LAMMPS} with parameters as described below. 
For the sake of clarity, 
we briefly reproduce below the key equations, 
but for a deeper understanding, 
the reader should refer to the original sources.

The GB particle shape is controlled by the ellipsoidal semiaxes $a_i$, $b_i$, and $c_i$ for
each particle $i$, which form the diagonal elements of
a ``shape'' matrix, $\vect{S}_i={\text{diag}(a_i,\,b_i,\,c_i)}$. 
The relative well depths $\varepsilon_{a,\,i}$, $\varepsilon_{b,\,i}$, and $\varepsilon_{c,\,i}$ 
for each particle $i$ give the matrix 
$\vect{E}_i={\text{diag}(\varepsilon_{a,\,i},\,\varepsilon_{b,\,i},\,\varepsilon_{c,\,i})}$.
We have the following relations:

\begin{equation}
\varepsilon_{a}=\sigma\cdot\frac{a}{b\cdot c};\,
\varepsilon_{b}=\sigma\cdot\frac{b}{a\cdot c};\,
\varepsilon_{c}=\sigma\cdot\frac{c}{a\cdot b}\,,
\end{equation}

\noindent where $a$, $b$, and $c$ control the shape of the ellipsoid in the side, 
face, and end-to-end dimensions, respectively.
More precisely, $a$, $b$, and $c$ are the
separations at which the attractive and repulsive terms in the
potential cancel when the molecules are in the side-by-side, face-to-face, 
and end-to-end  configuration, respectively.
On the other hand, $\varepsilon_{a}$,
$\varepsilon_{b}$, and $\varepsilon_{c}$, are the
well depths for the side-by-side, face-to-face, and end-to-end configurations, 
respectively.

The GB formula is,

\begin{equation}
\label{eqn:gb}
U_{\rm GB}=
U_{r}\cdot
\eta\cdot
\chi\,.
\end{equation}

Here, $U_{r}=4\varepsilon(\varrho^{12}-\varrho^{6})$,
with $\varrho=\dfrac{\sigma}{h_{ij}+\gamma\sigma}$,
controls the shifted distance-dependent interaction 
based on the distance of the closest approach between particles 
($h_{ij}(\vect{A}_{i},\,\vect{A}_{j},\,\vect{S}_{i},\,\vect{S}_{j},\,\vect{r}_{ij})$), where 
$\vect{A_{i}}$ and $\vect{A}_{j}$ are 
the transformation matrices 
from the simulation body-frame to the lab-frame and
$\gamma$ is the shift for potential minimum (here set to 1). 
Moreover, the interaction anisotropy is characterized by the 
terms $\eta(\vect{A}_{i},\,\vect{A}_{j},\,\vect{S}_{i},\,\vect{S}_{j},\,\nu)$ and 
$\chi(\vect{A}_{i},\,\vect{A}_{j},\,\vect{E}_{i},\,\vect{E}_{j},\,\hat{\vect{r}}_{ij},\,\mu)$ 
that control interaction strength based on the
particle shapes and relative well depths, respectively.
The parameters $\mu$ and $\nu$ 
are empirically determined exponents that can be tuned 
to adjust the potential. 

In this work, the liquid crystal units are modeled as uniaxial ellipsoids with 
a mass of $1.5\,m$, and GB parameters 
$\sigma=1$, $\varepsilon=1$, $a=b=\sigma$, $c=3\,\sigma$, 
$\varepsilon_{a}=\varepsilon_{b}=\varepsilon$, and 
$\varepsilon_{c}=0.2\,\varepsilon$.
With these parameters, we can define the aspect ratio, 
$\kappa\equiv c/a$ of the particles (in this case is 3)
and the energy anisotropy ratio, 
$\kappa'\equiv\varepsilon_{a}/\varepsilon_{c}$, 
in this case, set to 5.

From here, the GB potential has four adjustable parameters. 
In the compact notation of \citet{bates1999computer} 
these are ($\kappa$, $\kappa'$, $\mu$, $\nu$). 
The four parameters have been set to the original values 
\cite{gay1981modification} (3, 5, 2, 1) in several studies 
\cite{adams1987computer,emsley1992computer,
de1992dynamics,moreno2011effects}.
A second set of parameters, 
(3, 5, 1, 2), was studied 
by \citet{luckhurst1990computer}; here, the existence 
of smectic, nematic, and isotropic phases was demonstrated, 
while the behavior inside the phases was not studied in detail. 
This set of parameters was also implemented 
in side-chain liquid crystalline polymer studies 
\cite{ilnytskyi2007structure,ilnytskyi2012modelling}.
A third set of parameters that enhance the side-by-side 
and end-to-end interactions between the particles 
has been used 
by Berardi $et\,al.$ \cite{berardi1993monte,berardi2004computer}, 
(3, 5, 1, 3). 
This set has also been used by \citet{brown2009liquid} 
for studying liquid crystal nanodroplets in solution.
A final type of selected Gay--Berne potential 
that we are aware of is the one proposed 
by \citet{bates1999computer} and also implemented in 
\citet{margola2018comparison}, (4.4, 20, 1, 1). 
With this choice, the mesogens are closer 
to elongated molecules because the parametrization 
has larger anisotropic ratios than the other sets.
In this work, we have adopted the parametrization proposed by 
\citet{berardi1993monte}, (3, 5, 1, 3).
The values of the parameters used in the $U_{\text{nonbonded}}$ 
are given in the supplementary material.

\subsubsection{Bond and angle interactions}

\begin{figure}
	\centering
	\includegraphics[width=.7\textwidth]{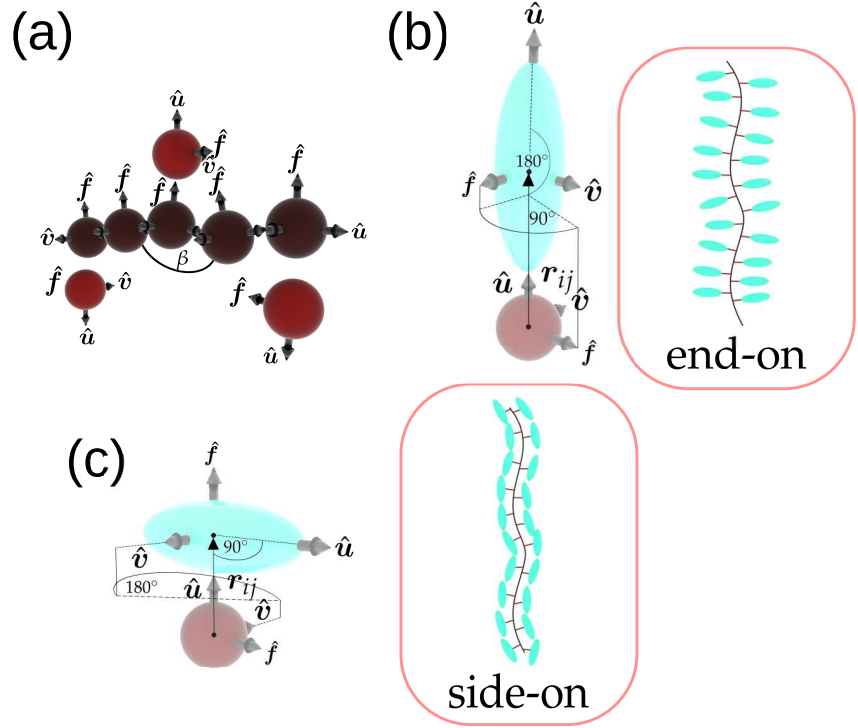}
	\caption{(a) Bonded interactions for the polymer backbone (dark scarlet) 
    and some spacer beads (scarlet) attached to it. 
	Here, $\beta$
	represents the bending angle between three adjacent atoms 
	and the $\hat{\vect{f}},\,\hat{\vect{v}},\,\hat{\vect{u}}$ unit vectors
	(arrows) 
	denote the orientation of each anisotropic site.	
	(b) Bonded interactions between the mesogenic unit (cyan ellipsoid)
	 attached in an end-on manner to a spacer bead (scarlet).
	(c) Bonded interactions between the mesogenic unit (cyan ellipsoid)
	 attached in a side-on manner to a spacer bead (scarlet).
	}
	\label{fig:intra}
\end{figure}

Since all sites in the model have both a position and an orientation, 
the bond and angle interactions in the SCLCP model 
can control the mesogenic type of attachment, type of tacticity, 
among other constraints that involve relative position 
and relative orientation of bonded sites.
To explain the intramolecular potentials, we separate them into 
potentials within the polymer backbone and the spacers, 
between the polymer backbone and spacers, and between 
the spacers and liquid crystals moieties (see Figure~\ref{fig:intra}). 
From now on, the notation employed for intramolecular potentials 
is the same as in \citet{lequieu20191cpn}.

Interactions within the polymer (main backbone and spacers) are 
represented by a twistable worm-like chain model 
\cite{chirico1994kinetics,brackley2014models,lequieu20191cpn}.
These flexible polymer chains consist of
$N_{b}$ beads, and the potential energy is given by

\begin{multline}
\label{eqn:}
U_{\text{bond}} + U_{\text{angle}} + U_{\text{twist}} + U_{\text{align}} \\
=\frac{k_{\text{b}}}{2} 
\sum_{i=1}^{N_{b}-1}(\ell_{i}-\ell_{\text{b},\,0})^{2}
+{k_{\beta}} 
\sum_{i=1}^{N_{b}-2}[1-\text{cos}(\beta_{i})]\\
+{k_{\omega}} 
\sum_{i=1}^{N_{b}-1}[1-\text{cos}(\omega_{i})]  
+{k_{\psi}} 
\sum_{i=1}^{N_{b}-1}[1-\text{cos}(\psi_{i})]\,,  
\end{multline}

\noindent where $\ell_{i}=|\vect{r}_{i+1}-\vect{r}_{i}|$ is the 
distance between adjacent sites, $k_{\text{b}}$ is the bond force constant, 
and $\ell_{\text{b},\,0}$ is the equilibrium bond length; 
$\text{cos}(\beta_{i})=
\hat{\vect{r}}_{(i)(i+1)}\cdot\hat{\vect{r}}_{(i+1)(i+2)}$
is the angle between adjacent bonds with angle force constant $k_{\beta}$;
$\text{cos}(\omega_{i})=\frac{\hat{\vect{f}}_{i+1}\cdot\hat{\vect{f}}_{i}
+\hat{\vect{v}}_{i+1}\cdot\hat{\vect{v}}_{i}}
{(1+\hat{\vect{u}}_{i}\cdot\hat{\vect{u}}_{i+1})}$ is the twist around 
the polymer backbone between adjacent sites and $k_{\omega}$ is 
the corresponding twist force constant. 
The last term, $U_{\text{align}}$, constrains $\hat{\vect{u}}_{i}$ 
of each site to be aligned with the bond vector, $\hat{\vect{r}}_{(i)(i+1)}$, 
where $\text{cos}(\psi_{i})=\hat{\vect{u}}_{i}\cdot\hat{\vect{r}}_{(i)(i+1)}$ 
and $k_{\psi}$ is the corresponding force constant. 
Note that through $k_{\beta}$ we can set the stiffness of the polymer. 

In the following, the bonded potentials regarding 
polymer main backbone (PB)--spacer (SC) interactions and 
spacer (SC)--liquid crystal (LC) 
interactions have a harmonic functional form, 
and the difference only lies in the value of the underlying parameters. 
We have adopted the notation 
$U(\ell_{ij};\,k,\,\ell_{0})=\frac{k}{2}(\ell_{ij}-\ell_{0})^{2}$ 
for the corresponding bonds, and 
$U(\hat{\vect{x}},\,\hat{\vect{y}};\,
k,\,\theta_{0})=\frac{k}{2}(\theta-\theta_{0})^{2}$ 
for the corresponding angles, 
where $\ell_{ij}=|\vect{r}_{ij}|$, 
$\text{cos}(\theta)=\hat{\vect{x}}\cdot\hat{\vect{y}}$, and 
$k$, $\ell_{0}$, and $\theta_{0}$ are the corresponding 
force constant, equilibrium bond length, and equilibrium angle 
\cite{lequieu20191cpn}. 

To attach the spacers to the main backbone while ensuring their syndiotacticity, 
we have adopted a bond potential and an orientational potential as follows

\begin{multline}
U_{\text{bond}} + U_{\text{orient}}
=U_{\text{b}}(\ell_{\text{PB},\,i-\text{SC},\,j};
\,k_{\text{b}}^{N_{\text{SC}}},\,\ell_{\text{b},\,0})\\
+U_{\text{o}}(\hat{\vect{r}}_{ij},\,\hat{\vect{v}}_{\text{PB},\,i};\,k_{\alpha}^{N_{\text{SC}}},\,90^{\circ})
+U_{\text{o}}(\hat{\vect{r}}_{ij},\,\hat{\vect{v}}_{\text{SC},\,j};\,k_{\alpha}^{N_{\text{SC}}},\,90^{\circ})
+U_{\text{o}}(\hat{\vect{v}}_{\text{PB},\,i},\,\hat{\vect{v}}_{\text{SC},\,j};\,k_{\alpha}^{N_{\text{SC}}},\,180^{\circ})\\
+U_{\text{o}}(\hat{\vect{r}}_{ij},\,\hat{\vect{u}}_{\text{PB},\,i};\,k_{\alpha}^{N_{\text{SC}}},\,90^{\circ})
+U_{\text{o}}(\hat{\vect{r}}_{ij},\,\hat{\vect{u}}_{\text{SC},\,j};\,k_{\alpha}^{N_{\text{SC}}},\,0^{\circ})
+U_{\text{o}}(\hat{\vect{u}}_{\text{PB},\,i},\,\hat{\vect{u}}_{\text{SC},\,j};\,k_{\alpha}^{N_{\text{SC}}},\,90^{\circ})\,,
\end{multline} 

\noindent where ${N_{\text{SC}}}$ is the number of spacers, and the angles  
are the result from the dot product between the corresponding 
unit orientational vectors specified.
From here it can be deduced that if one is interested in 
isotactic or atactic configurations, it would suffice to modify 
or eliminate the $U_{\text{orient}}$ potential, respectively.

Finally, we have employed a bonding potential to attach the 
terminal mesogenic group that represents a liquid crystal unit 
to the spacers and torsional potentials that 
prevent the ellipsoids from unphysical rotations
with respect to the rest of the chain and also set the liquid crystal 
type of attachment; end-on side-chain (EOSC)
and side-on side-chain (SOSC) configurations.

For the EOSC configuration, the potential is given by

\begin{multline}
U_{\text{bond}} + U_{\text{orient}}\\
=U_{\text{b}}(\ell_{\text{SC},\,i-\text{LC},\,i+1};
\,k_{\text{b,\,EOSC}}^{N_{\text{SC}}},\,\ell_{{\rm b},\,0,\,{\rm EOSC}})
+U_{\text{o}}(\hat{\vect{f}}_{\text{SC},\,i},\,\hat{\vect{f}}_{\text{LC},\,i+1};\,k_{\xi}^{N_{\text{SC}}},\,90^{\circ})\\
+U_{\text{o}}(\hat{\vect{r}}_{(i)(i+1)},\,\hat{\vect{u}}_{\text{SC},\,i};\,k_{\xi}^{N_{\text{SC}}},\,0^{\circ})
+U_{\text{o}}(\hat{\vect{r}}_{(i)(i+1)},\,\hat{\vect{u}}_{\text{LC},\,i+1};\,k_{\xi}^{N_{\text{SC}}},\,0^{\circ})\,.
\end{multline}

\noindent and for the SOSC configuration, the potential is given by

\begin{multline}
U_{\text{bond}} + U_{\text{orient}}\\
=U_{\text{b}}(\ell_{\text{SC},\,i-\text{LC},\,i+1};
\,k_{\text{b,\,SOSC}}^{N_{\text{SC}}},\,\ell_{{\rm b},\,0,\,{\rm SOSC}})
+U_{\text{o}}(\hat{\vect{v}}_{\text{SC},\,i},\,\hat{\vect{v}}_{\text{LC},\,i+1};\,k_{\xi}^{N_{\text{SC}}},\,180^{\circ})\\
+U_{\text{o}}(\hat{\vect{r}}_{(i)(i+1)},\,\hat{\vect{u}}_{\text{SC},\,i};\,k_{\xi}^{N_{\text{SC}}},\,0^{\circ})
+U_{\text{o}}(\hat{\vect{r}}_{(i)(i+1)},\,\hat{\vect{u}}_{\text{LC},\,i+1};\,k_{\xi}^{N_{\text{SC}}},\,90^{\circ})\,,
\end{multline} 

Quaternion-based methods for time integration of 
anisotropic sites and thermostatting/barostatting of their 
rotational degrees of freedom are taken into account
in this work. 
All the parameter values used for $U_{\rm bonded}$  
are given in the supplementary material.  

\subsection{Simulation methodology}
\label{subsec:methodology}

The SCLCP molecules are grown from random positions inside 
the computational box by treating each molecule 
as a 3D random walk.
The resultant configuration consists of overlapping 
SCLCP molecules.
The MD simulations are performed using 
the open-source LAMMPS package \cite{LAMMPS}. 
The equations of motion are propagated via 
the velocity-Verlet algorithm with a time step of
$0.002\,\tau$, where $\tau=\sigma\sqrt{m/\varepsilon}$.
Periodic boundary conditions are applied 
in all three uncoupled dimensions of the box.
The first equilibration steps are performed with 
a soft repulsive potential (slow push-off method) 
which allows SCLCPs to pass through each other, 
introducing excluded volume interactions incrementally.
After equilibrating the melt with the slow push-off method, 
systems are equilibrated 
in the isothermal-isobaric (NPT) ensemble using a Langevin thermostat
and a Berendsen barostat
with the full potentials for at least 
3 times the longest end-to-end vector relaxation time at 
a temperature $T=8.0\,\varepsilon/k_{\text B}$ and
pressure $P=1.0\,\varepsilon{/\sigma}^{3}$ 
(see Subsec.~\ref{subsec:equilibration})
to obtain an isotropic melt
with no memory of the initial spatial arrangement.
The damping parameters for the thermostat and barostat 
are set to $0.5\,\tau$ and $5\,\tau$, respectively. 

To study mesophase behavior, 
we gradually cool down 
the systems from equilibrated isotropic melts at
$T=8.0\,\varepsilon/k_{\text B}$ and 
$P=1.0\,\varepsilon/{\sigma}^{3}$ until 
crystalline order is reached as explained below.

In this work, we study polymers with purely end-on and 
purely side-on LC groups separately.
We consider different variations of architectural 
parameters of the SCLCP model, namely; 
two backbone lengths, 
three spacer lengths, and 
three polymer stiffnesses 
(this is controlled by the angle force constant, $k_{\beta}$).
This gives rise to 12 different systems where 
the total number of particles 
is kept constant for all simulated cases.
Here, the notation employed for the systems is as follows [A:B:C], 
where A is the number of backbone beads, 
B is the number of spacer beads, and C is the angle force constant 
in dimensionless units.
Table~\ref{tab:sclcpcases} shows every case simulated in this work.
Note that only for comparison purposes we have also simulated 
a melt of linear polymer chains [100:0:1.5] without LC units attached.

The mean properties are represented 
by statistical averages over all of the SCLCPs 
of the system configurations evenly taken from 
the last steps of a particular simulation. 

\begin{table}
\caption{Side-chain liquid crystal polymer systems studied with 
their corresponding notation that reflects their architectural properties. 
Each system is replicated and studied  
for both end-on side-chain (EOSC) and side-on side-chain (SOSC)
liquid crystal polymers.}
\label{tab:sclcpcases}
\begin{tabular}{rcc}
\hline
\hline
System & Number of SCLCP molecules ($N_{\rm m}$) & Number of particles ($N_{\rm p}$) \\ 
\hline
[44:6:1.5]  & 636 & 125,928\\
 \hline 
 [100:2:1.5] & 504 & 126,000 \\ 
 \hline 
 [100:4:1.5] & 360 & 126,000\\
 \hline
 [100:6:0.0] & 280 & 126,000\\  
 \hline 
 [100:6:1.5] & 280 & 126,000 \\  
 \hline 
 [100:6:5.0] & 280 & 126,000\\ 
 \hline
\hline
\end{tabular}
\end{table}

\subsubsection{Cooling simulations}

Here, a cooling rate of $5\times 10^{-5}\,(\varepsilon/k_{\rm B})/\tau$ 
is employed. 
The studied temperatures are: 
8.0, 7.0, 6.0, 5.0, 4.0, 3.0, 2.5, 2.0, 1.5, and 1.0.
From the starting relaxed isotropic configuration of each system 
at $T=8.0\,\varepsilon/k_{\text B}$ and 
$P=1.0\,\varepsilon/{\sigma}^{3}$, the samples are cooled
using the specified cooling rate in NVT ensemble, 
and at each temperature studied the system is relaxed 
in NPT ensemble at $P=1.0\,\varepsilon/{\sigma}^{3}$ 
during $4,000\,\tau$.
The natural evolution of these systems is 
to display local ordering of the liquid crystals 
as the temperature decreases. 
To obtain global and positional ordering of the LC layers 
one could apply an external alignment field that induces a preferential orientation 
of LC molecules as explained below. 
Therefore, we apply this external field to the main axis 
of LC moieties in a particular direction during 
the cooling simulations.
This alignment field is switched off in the relaxation simulations 
at the specific studied temperatures.
When the field is switched off at high temperatures, 
the global ordering vanishes, 
but this changes when moving to lower temperatures, 
where the structure acquires global and positional ordering.

\subsubsection{External alignment field}

To induce artificial alignment of liquid crystals 
with global ordering and to avoid the 
long simulation times required for mesophase formation, 
we have implemented an external alignment field 
in LAMMPS \cite{LAMMPS} that applies a torque 
to the orientation of a particle, in this particular case, 
to the main axis of the uniaxial ellipsoids 
representing liquid crystals with respect 
to a fixed axial direction. 

This external torque applied to the molecules mimics an external magnetic field applied 
to liquid crystals, which are employed in experiments, 
to align them parallel to the field, 
thus obtaining global ordering \cite{wang2022multi}.
Similar potentials used in previous works 
\cite{stimson2005molecular,ilnytskyi2012modelling} 
have shown that both the external field and 
its strength affect the structures formed in the microphase.

In particular, the applied torque is as follows

\begin{eqnarray}
\vect{\tau}_{\text{align}}=\left\{
 \begin{array}{lr} 
k_{\text{align}}\hat{\vect{u}}_{i}\times\hat{\vect{e}} & \hspace{1cm}{\rm if}\,\hat{\vect{u}}_{i}\cdot\hat{\vect{e}}>0 \\
& \\
-k_{\text{align}}\hat{\vect{u}}_{i}\times\hat{\vect{e}} & \hspace{1cm}{\rm if}\,\hat{\vect{u}}_{i}\cdot\hat{\vect{e}}<0
\end{array}
\right.
\end{eqnarray}

\noindent were $k_{\text{align}}$ is the force constant that determines 
how aggressive is the external alignment field, and $\hat{\vect{e}}$ 
is the unit directional vector.
The direction of $\hat{\vect{e}}$ can be set in any direction of a 
fixed 3D space. 
This torque allows us for a parallel alignment 
of the LC moieties with respect to the field, regardless of whether 
their main axis ($\hat{\vect{u}}_{i}$)
is pointing in the opposite direction as $\hat{\vect{e}}$.
Here, $k_{\text{align}}$ is set to $45\,k_{\rm B}T$.

\section{RESULTS AND DISCUSSION}
\label{sec:results}

\subsection{Equilibration of the systems}
\label{subsec:equilibration}

\begin{figure}
	\centering
	\includegraphics[width=1\textwidth]{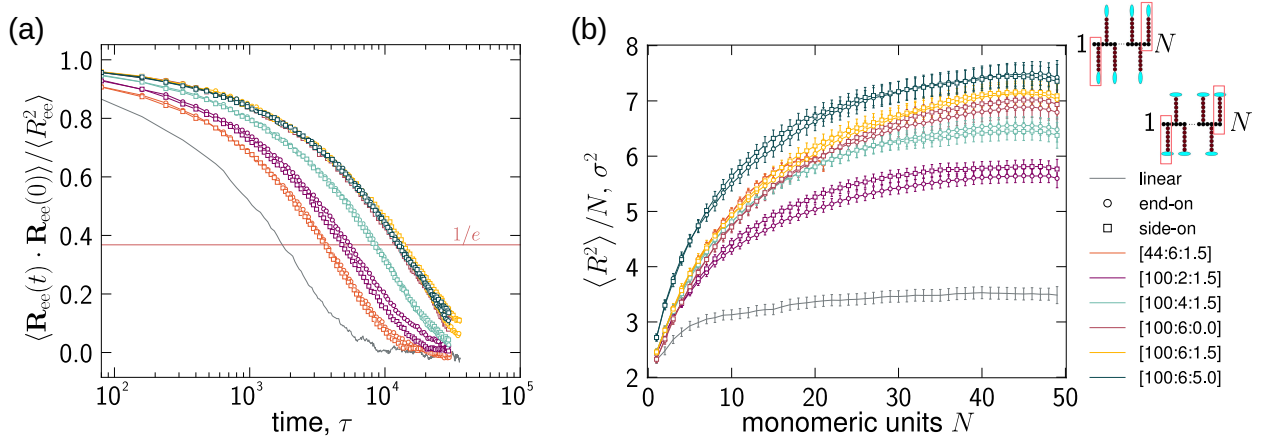}
	\caption{(a) End-to-end vector autocorrelation function 
	($\rm ACF_{\rm ee}$) 
	for all the melt systems studied.
	(b) Mean-squared internal distances $\langle R^{2}\rangle/N$
	as a function of monomeric units $N$ for all the equilibrated melt systems 
	at $T=8.0\,\varepsilon/k_{\text B}$ 
	and $P=1.0\,\varepsilon/{\sigma}^{3}$.
	Simulations for end-on side-chain (EOSC) and side-on side-chain (SOSC)
	configurations are shown for each case with different architectural 
	parameters.
	Results for a linear polymer melt modeled with the same 
	SCLCP backbone potential 
	are included for comparison in (a) and (b). 
	Schemes of how monomeric units are defined in SCLCP molecules 
	are depicted in (b).
	}
	\label{fig:endtoendacf}
\end{figure}

We calculate chain observables and examine
their dependence on architectural parameters.
All the measurements are performed on melt systems 
after the equilibration process at $T=8.0\,\varepsilon/k_{\text B}$ 
and $P=1.0\,\varepsilon/{\sigma}^{3}$.
Firstly, we analyze the backbone's end-to-end vector relaxation.
Specifically, we calculate 
the end-to-end vector autocorrelation function ($\rm ACF_{ee}$) 
defined as 
$\dfrac{\langle{\vect{R}}_{\text{ee}}(t)
\cdot{\vect{R}}_{\text{ee}}(0)\rangle}
{\langle R_{\text{ee}}^{2}\rangle}$,
where ${\vect{R}}_{\text{ee}}$ is the vector from the 
first backbone bead to the last backbone bead.
The function was calculated for time blocks of $30,000\,\tau$ 
with starting times spaced $2,000\,\tau$ apart. 
These block-averaged functions were then averaged over 
15 trials with different starting configurations of the SCLCP melt systems.
Figure~\ref{fig:endtoendacf}a shows 
the mean backbone's end-to-end vector relaxation time for the systems 
with different backbone lengths, spacer lengths, 
angle force constants, and mesogenic types of attachment 
(end-on and side-on configurations).
To ensure proper relaxation of SCLCP systems,
equilibration times are at least 3 times 
the longest end-to-end vector relaxation time 
(the intersection at $1/e$).
Comparing the SCLCP systems with the linear polymer melt system, 
we found that the presence of side chains (spacers and LC moieties) slows down the relaxation 
of the former by one order 
of magnitude on average, as would be intuitively expected.  
Specifically, for SCLCP systems, the end-to-end vector relaxation time 
is found to be dependent on the backbone length and spacer length.
This effect is expected due to the presence of polymer entanglements
in systems composed of long polymer chains, even when our systems 
should be only slightly entangled for chains composed
of 100 beads. 
Our results also show that the mesogen type of attachment 
and angle force constant ($k_{\beta}$) have a minimal or no effect on 
the end-to-end autocorrelation time, 
at least for the angle force constants studied here 
since it is well studied that entanglement properties vary with chain stiffness 
\cite{bobbili2020simulation}.

Secondly, for the mean-squared internal distance curves 
shown in Figure~\ref{fig:endtoendacf}b, 
each $\langle{R}^{2}\rangle/N$ point (with $N$ the number of monomeric units),
where a monomer in the SCLCP model is defined as 
two consecutive backbone beads and the corresponding 
side chain that is attached to one of them,
is calculated by averaging over all of the chains of 10
melt configurations, spaced $9,000\,\tau$ from each other and from
the last $90,000\,\tau$ of simulation.
In all cases, 
a plateau in the $\langle R^{2}\rangle/N\,vs\,N$ curve is obtained.
The mean-squared internal distances are strictly correlated to
chain structure and have shown to be non-dependent on the 
mesogenic type of attachment.
It can be observed that within the
statistical error bars (standard error), there are no significant differences between 
EOSC and SOSC configurations.
On the other hand, the $\langle R^{2}\rangle/N$ 
observable is strongly dependent on 
the spacer length, 
due to steric effects, 
and on the angle force constant, 
being greater for higher values of $k_{\beta}$, 
which is intuitive since one would expect 
stiffer chains to be more elongated. 
Finally, a trend can be observed for different backbone 
lengths, specifically if one compares between [44:6:1.5] 
and [100:6:1.5] cases.
Here, for linear polymer chains of a particular chemistry, 
the value of the plateau 
should be the same for the same temperature regardless 
of backbone length (if normalized by molecular weight), however, for SCLCP molecules, 
having longer backbone chains means more side chains attached, 
which induces a higher excluded volume due to the steric effect, 
and these are more effective at keeping other chains away. \cite{bobbili2020simulation}.
This effect can be easily observed also if one compares 
the different plateau values for the [100:6:1.5] and the 
melt composed of linear polymer chains [100:0:1.5] 
that share the same number of backbone beads.

\subsection{Structure of mesophases}
\label{subsec:structure-mesophases}

After corroborating equilibration and relaxation 
at high temperatures,
we next proceed to investigate the structure of mesophases 
at different thermodynamic states.
In doing that, we slowly cool down the systems to lower temperatures, 
while, as explained above, an external alignment field 
is applied intermittently during the cooling process 
to ensure alignment occurs in a given direction, 
avoiding multiple crystalline domains with different orientations. 

\begin{figure}
	\centering
	\includegraphics[width=.9\textwidth]{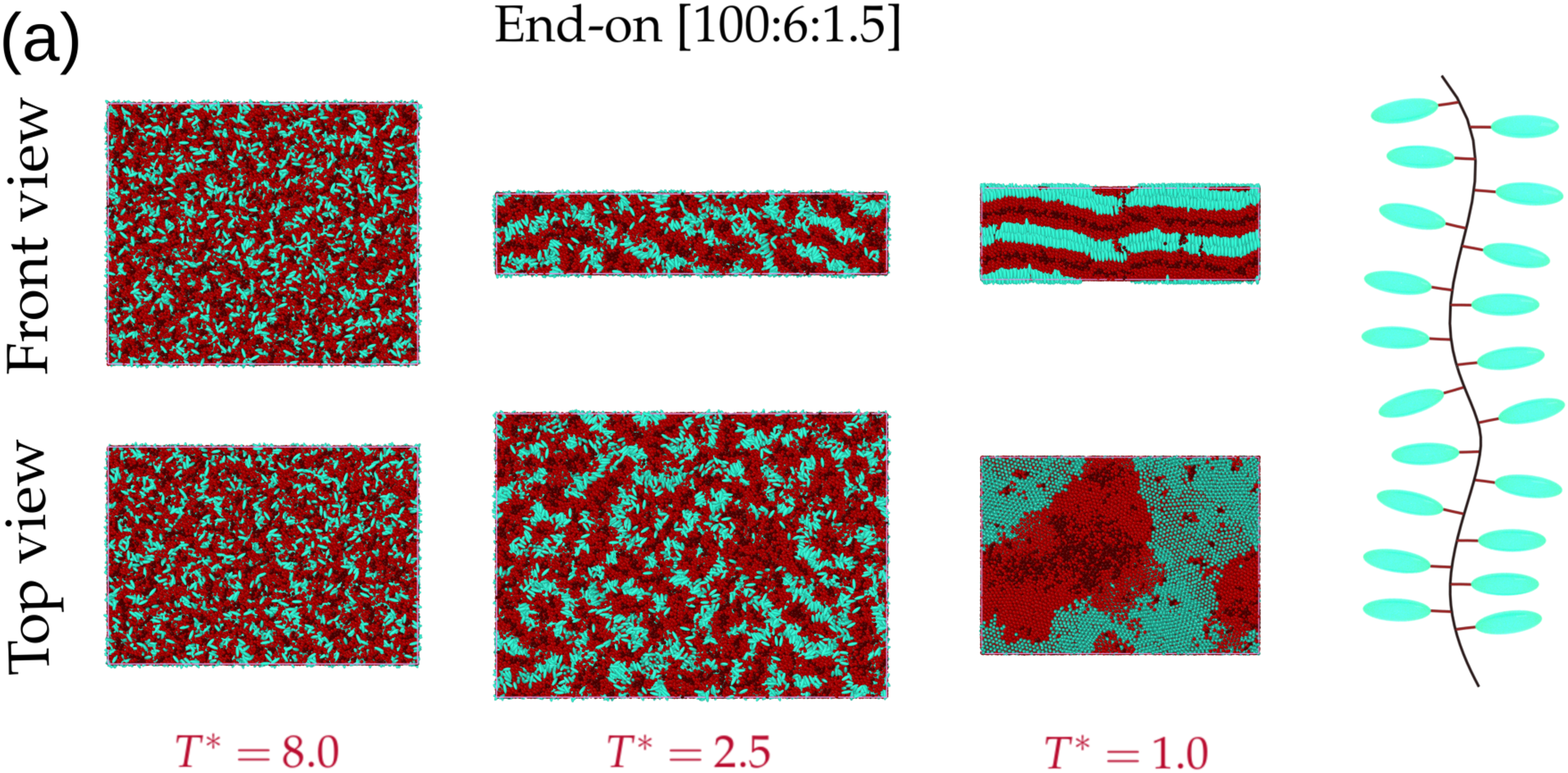}\\
	\includegraphics[width=.9\textwidth]{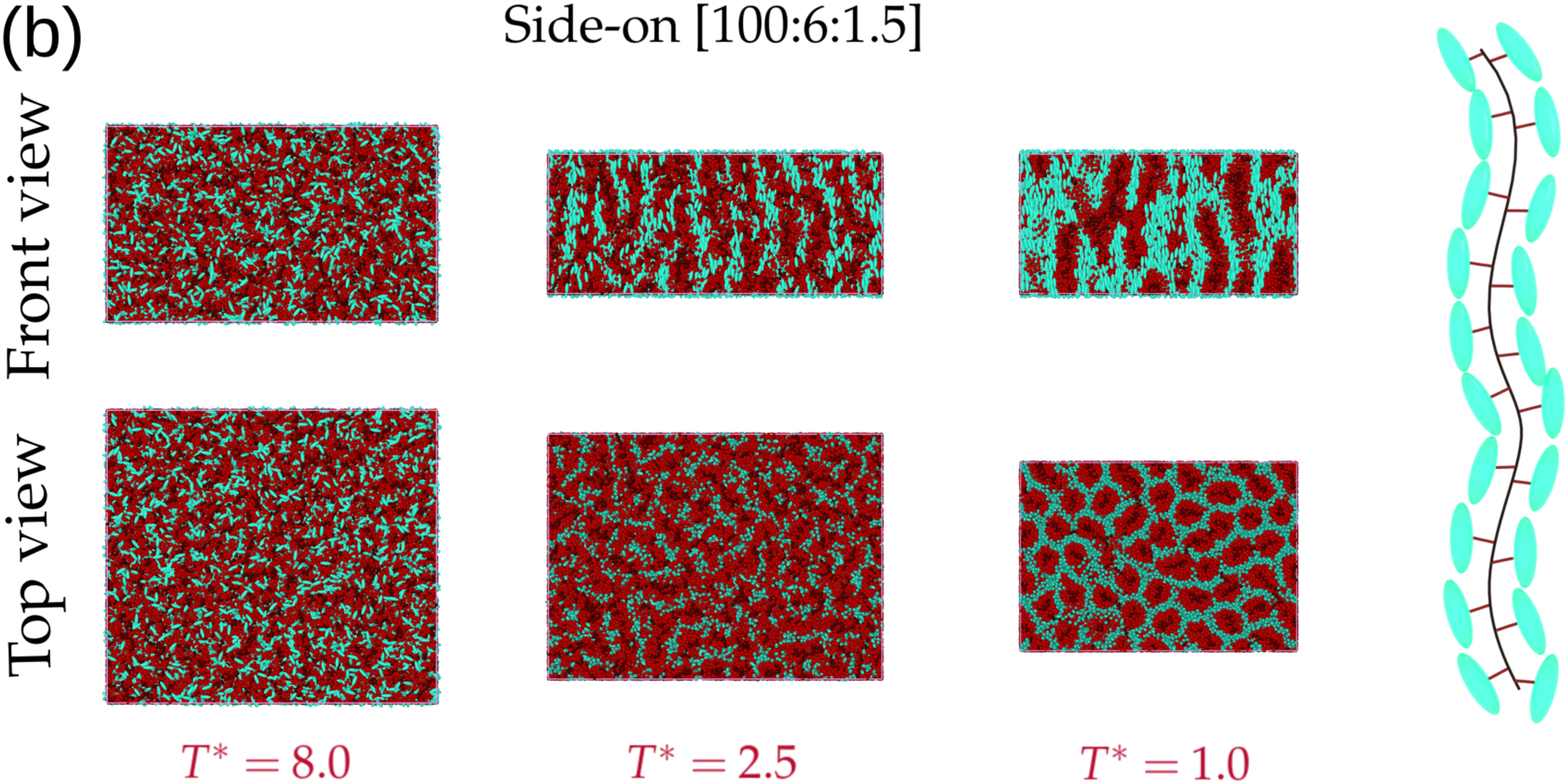}
	\caption{Snapshots of SCLCP [100:6:1.5] systems (top and front views) 
	from equilibrium simulations at different thermodynamic states 
	for (a) end-on side-chain (EOSC) configuration and 
	(b) side-on side-chain (SOSC) configuration.
	From left to right, mesophases from isotropic 
	to crystalline order can be observed,
	where cyan represents liquid crystals, scarlet represents
	spacers, and dark scarlet represents polymer backbones.
	}
	\label{fig:configurations_box}
\end{figure}

Through system visualizations using the software OVITO 
\cite{stukowski2009visualization}, 
we corroborate that SCLCP systems with only one type of LC moiety 
indeed form well-organized structures at low temperatures, 
as shown in Figure~\ref{fig:configurations_box} 
that remain stable even after the alignment field 
is switched off.

Interestingly, as can be seen in Figure~\ref{fig:configurations_box} 
for [100:6:1.5] systems, 
at low temperatures, purely end-on and side-on systems give rise to different morphologies. Figure~\ref{fig:configurations_box}a shows 
top and front views of the mesophases formed for the EOSC [100:6:1.5] system 
at 3 different temperatures from equilibrium runs 
(here, the external alignment field is switched off): 
at $T=8.0\,\varepsilon/k_{\rm B}$, the structure is 
completely isotropic without any kind of global or local ordering; 
at $T=2.5\,\varepsilon/k_{\rm B}$, it is not possible to appreciate
a global ordering of the mesogens but a local ordering can be observed, 
and also an elongation of the system in the direction perpendicular to which 
the alignment field is applied during cooling simulations;
at $T=1.0\,\varepsilon/k_{\rm B}$, it can be seen how the system 
slightly shrinks in the elongated direction due to decreased temperature
(see Figure~\ref{fig:configurations_box}a, middle and right snapshots), 
and a global ordering can be observed.
In particular, from the front view, it can be observed that 
liquid crystals, polymer backbones, and the spacers 
all display a global ordering, with the mesogens arranged in a bilayer (smectic B-like mesophase).
Also, from the snapshot at $T=1.0\,\varepsilon/k_{\rm B}$ 
we can conclude that the elongation is favored in the direction parallel to the 
polymer backbones, thus perpendicular to the LC director, 
in line with experimental results \cite{xu2021random}.
Moreover, the LC bilayer is slightly tilted, and some defects can be appreciated 
as in real networks \cite{kluppel1993characterization,ren2008poisson} 
and previous MD simulations \cite{ilnytskyi2012modelling}.
From the top view at low temperatures, the LCs adopt 
a hexagonal lattice, which indicates crystalline order. 
The smectic B-like phase displayed by end-on side-chain systems at low temperatures 
is in good agreement with what was found in previous works 
for smaller systems \cite{ilnytskyi2012modelling,stimson2005molecular}.
On the other hand, a similar behavior can be observed for side-on side-chain systems 
at high temperatures, that is, isotropic ordering 
without preferential alignment. 
However, unlike EOSC systems, at low temperatures the SOSC systems 
exhibit nematic ordering in one direction due to how the LCs are attached 
to the spacer, as shown in the front view of 
Figure~\ref{fig:configurations_box}b. 
In the top view of Figure~\ref{fig:configurations_box}b, 
it can be seen that LCs surround 
the polymer backbones. 
In this way, a remarkably hexagonal structure is formed,
which is possibly the most energetically stable configuration in this 
thermodynamic state.
The elongation of the system, in this case, is also in the direction perpendicular 
to the LC director but much less pronounced than in the EOSC case.

Snapshots that include all studied systems 
with different variations of the architectural parameters 
are shown in the supplementary material. 
Among the most notable variations, 
we find that for systems with short spacers ([100:2:1.5]), 
very high elongations occur in different directions 
depending on the LC configurations (either end-on or side-on), 
which we relate to the great coupling between LCs and the polymer backbones. 
Regarding polymer stiffness, we find that the most flexible polymer 
(EOSC [100:6:0.0] system)
forms a very well-ordered smectic B-like structure, 
where three LC bilayers can be observed inside the box 
and the same high ordering in a hexagonal structure 
for SOSC [100:6:0.0] systems; 
while the stiffest polymer systems 
in the EOSC [100:6:5.0] configuration 
shows only monolayers of LC units and a more elongated structure 
in the opposite direction of the LC director as expected and 
for the SOSC [100:6:5.0] configuration the ordering 
at low temperatures is not so remarkable.

\begin{figure}
	\centering
	\includegraphics[width=1\textwidth]{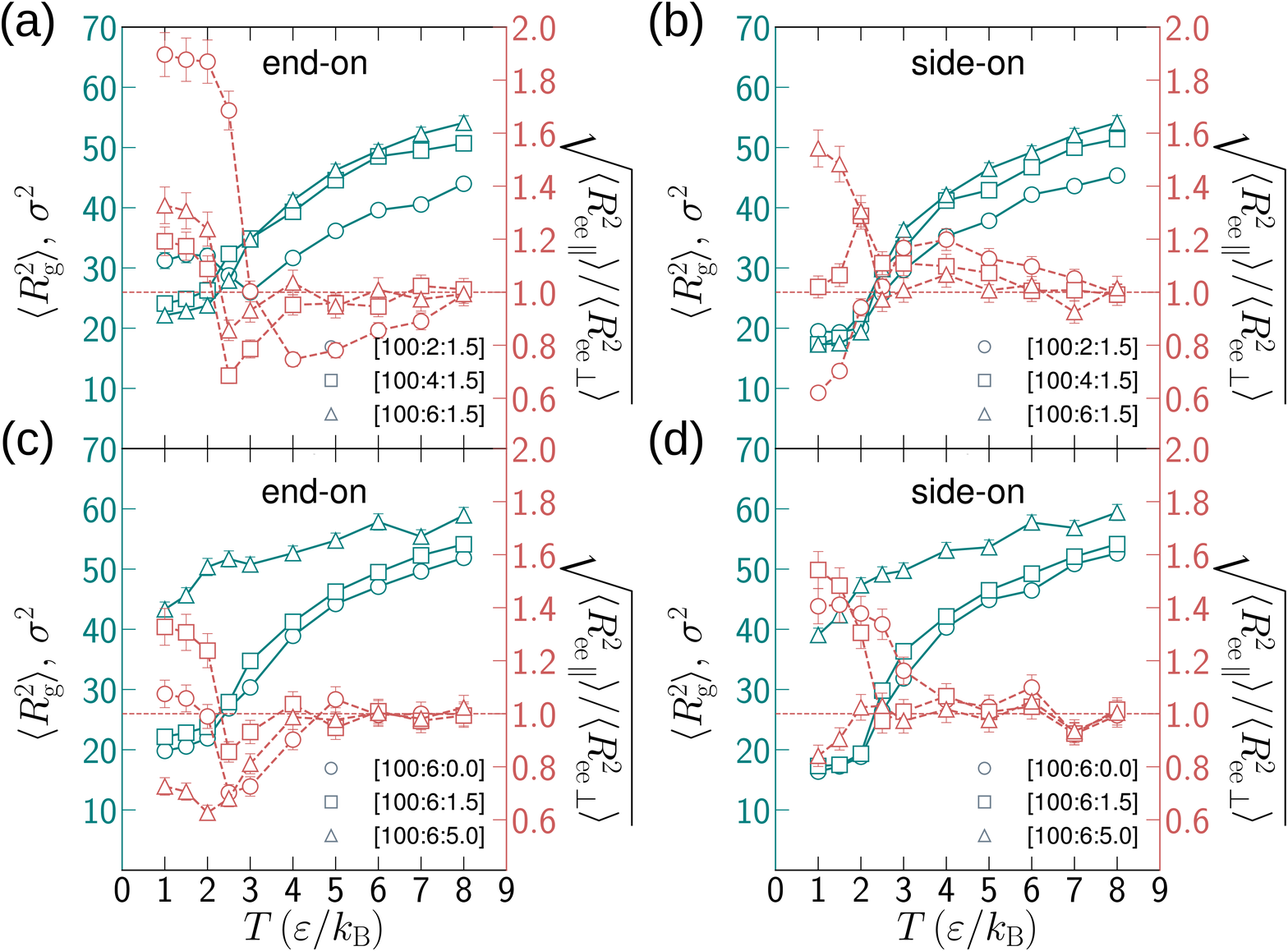}
	\caption{Mean-square radius of gyration of the backbone $\langle R_{\rm g}^{2}\rangle$ 
        (turquoise and continuous lines) 
	and the anisotropic ratio 
        $\sqrt{\langle R_{\rm ee\,\parallel}^{2}\rangle/\langle R_{\rm ee\,\perp}^{2}\rangle}$
        (red and dashed lines) for different systems
        at different temperatures from 
	equilibrium simulations.
        The standard errors are propagated in the latter and included in all the measurements.
	The effect of architecture parameters on $\langle R_{\rm g}^{2}\rangle$ 
        and $\sqrt{\langle R_{\rm ee\,\parallel}^{2}\rangle/\langle R_{\rm ee\,\perp}^{2}\rangle}$
        as a function of temperature
	is shown for different spacer lengths in
        a) for end-on side-chain (EOSC) configurations and in
        b) for side-on side-chain (SOSC) configurations, and
	for different angle force constants ($k_{\beta}$) in
        c) for end-on side-chain (EOSC) configurations and in
        d) for side-on side-chain (SOSC) configurations.  
	The isotropic value ($1$) for 
        $\sqrt{\langle R_{\rm ee\,\parallel}^{2}\rangle/\langle R_{\rm ee\,\perp}^{2}\rangle}$ 
        is depicted in red as a guide for the eye.
	}
	\label{fig:rgyration-ree}
\end{figure}

We calculate the mean-square radius of gyration of the polymer backbone 
for all the systems studied at different 
temperatures.
As shown in Figure~\ref{fig:rgyration-ree} the $\langle R_{\rm g}^{2}\rangle$ 
decreases for all the systems as the temperature decreases and reaches a 
plateau for the crystalline order at low temperatures.
If LC types of attachments are compared (left side against the right side in 
Figure~\ref{fig:rgyration-ree}) we observe only a slight difference between 
end-on and side-on configurations.
In some cases, at low temperatures, the end-on configurations display higher values than 
side-on configurations.

SCLCP systems with different backbones lengths are compared and as expected 
the [100:6:1.5] systems (with longer backbones) display higher values than the [44:6:1.5] system.
By varying the spacer length, a clear trend can be observed in 
Figures~\ref{fig:rgyration-ree}a,b;
at high temperatures, due to a steric effect, the $\langle R_{\rm g}^{2}\rangle$ 
is larger for longer spacers. 
This trend is maintained throughout the entire temperature range, 
the only exception is the EOSC [100:2:1.5] at low temperatures, 
which again could be due to an artifact for the [100:2:1.5] systems at low temperatures.
On the other hand, Figures~\ref{fig:rgyration-ree}c,d show 
a greater mean square radius of gyration for the stiffest polymer chain, 
which even at low temperatures holds a similar value to those displayed at high temperatures 
for both end-on and side-on configurations.

We also calculate the anisotropic ratio defined as 
$\sqrt{\langle R_{\rm ee\,\parallel}^{2}\rangle/\langle R_{\rm ee\,\perp}^{2}\rangle}$, 
where $\langle R_{\rm ee\,\parallel}^{2}\rangle$ is the mean square
end-to-end distance vector in the direction parallel to the
alignment field, and $\langle R_{\rm ee\,\perp}^{2}\rangle$
is the mean square end-to-end distance in the perpendicular
direction to the alignment field.

\citet{wang2010theory} employed this parameter to characterize their model
for SCLCPs.
They modeled the interaction between the polymer backbone and the LC side groups 
through two coupling effects: the global nematic coupling and the local
hinge effect. 
The global coupling involves the interaction of a
backbone segment with the average nematic field produced by all
LC side groups in its vicinity whereas the hinge effect involves the
instantaneous values of the orientation vectors of the chain
segment and its own attached LC side group, which is purely
an intrachain effect \cite{wang2010theory}.
Moreover, they neglected the volume occupied by the spacer in their model, 
but its effect in terms of interactions is implicitly included.
The base case that they employed consisted of one side chain per Kuhn segment,
but more graft densities are explored.
They found a value higher than 1 for all the cases studied for the side-on configurations, which 
means a prolate conformation. 
By decreasing grafting density, the anisotropic ratio decreased (becoming less prolate), 
but remained higher than 1.
On the other hand, the conformation for the end-on SCLCPs
displayed both oblate and prolate conformations depending on the graft density, 
since the two coupling effects are reported to be competitive in that case. 
Therefore for the base case, they obtained values smaller than 1 for the anisotropic ratio
$\sqrt{\langle R_{\rm ee\,\parallel}^{2}\rangle/\langle R_{\rm ee\,\perp}^{2}\rangle}$ 
which means oblate conformations ($\langle R_{\rm ee\,\perp}^{2}\rangle>\langle R_{\rm ee\,\parallel}^{2}\rangle$),
but for decreased grafting density values higher than 1 were displayed.

Figure~\ref{fig:rgyration-ree} shows the anisotropic ratio
$\sqrt{\langle R_{\rm ee\,\parallel}^{2}\rangle/\langle R_{\rm ee\,\perp}^{2}\rangle}$ 
the systems studied.
From $T=8\,\varepsilon/k_{\rm B}$ to $T=4\,\varepsilon/k_{\rm B}$
is safe to say that all the systems display the isotropic value of one 
with the only exception of the [100:2:1.5] systems due to the higher 
LC-backbone coupling.
That means even at high temperatures the EOSC [100:2:1.5] (Figure~\ref{fig:rgyration-ree}a)
display oblate conformation (anisotropic ratio smaller than 1) and 
the SOSC [100:2:1.5] (Figure~\ref{fig:rgyration-ree}b)
display prolate conformation (anisotropic ratio higher than 1).
At lower temperatures, these two systems completely change their conformation, 
which may be due to simulation artifacts or probably 
some molecular effects in the crystalline regime.
If we compare Figure~\ref{fig:rgyration-ree}c,d where different chain stiffness
are studied for end-on and side-on configurations, respectively, 
two clear trends can be observed;
(i) the anisotropic ratio of the end-on SCLCP systems tends to decrease at temperatures lower 
than $4\,\varepsilon/k_{\rm B}$ up to $T=2.5\,\varepsilon/k_{\rm B}$ close to the isotropic to 
liquid crystal transition.
In this range the anisotropic ratio adopts values smaller than 1, therefore the chains adopt an oblate 
conformation in agreement with the theoretical model of \citet{wang2010theory} and 
experimental results \cite{xu2021random}.
At temperatures lower than $2.5\,\varepsilon/k_{\rm B}$, in the crystalline regime
the anisotropic parameter display values higher than 1, which means that the chains adopt a 
prolate conformation.
We infer that this effect could be to some molecular coupling at very low temperatures.
The same trend is followed for the EOSC [100:4:1.5].
(ii) The anisotropic ratio of the side-in SCLCP systems tends to hold the isotropic 
value of 1 up to $T=2\,\varepsilon/k_{\rm B}$ since the transition temperatures are lower for these 
systems when compared with the end-on side-chain systems.
At lower temperatures in the crystalline regime, the anisotropic ratio tends to adopt values higher than 1, 
therefore the chains display prolate conformations (see Figures~\ref{fig:rgyration-ree}c,d) 
in agreement with experimental results \cite{xu2021random} and theory \cite{wang2010theory}.
The SOSC [100:6:5.0] system represents an exception here along with the SOSC [100:2:1.5] system.
The former could be due to the stiffness affecting the rotational 
coupling between the spherical and aspherical beads of the model.

\subsubsection{Orientational order parameter}

To quantify the overall degree of molecular ordering, 
an instantaneous alignment second-rank tensor
is defined as

\begin{equation}
Q_{\alpha\beta}=
\frac{1}{N_{\rm LC}}\sum_{i=1}^{N_{\rm LC}}\frac{3}{2}
\left[(\hat{\vect{u}}_{i})_{\alpha}(\hat{\vect{u}}_{i})_{\beta}
-\frac{1}{3}\delta_{\alpha\beta}
\right]\,,
\end{equation}

\noindent where $N_{\rm LC}$ is the number of LC units in the system, 
$\delta_{\alpha\beta}$ is the Kronecker delta function, and 
$(\hat{\vect{u}}_{i})_{\alpha}$ are the Cartesian coordinates 
($\alpha=x,\,y,\,z$) of the main axis of 
the $i$th mesogen ($i=1,\,2,\,...,\,N_{\rm LC}$)
(see reference \cite{mottram2014introduction} for more details
on this calculation). 
The time average of the largest eigenvalue
of the alignment tensor $Q_{\alpha\beta}$ is equal to the orientational
order parameter $S_{2}$ 
(which is in fact the second-order Legendre polynomial, 
$P_{2}({\rm cos}\,\theta_{\rm m})$, where $\theta_{\rm m}$ 
is the angle between a LC particle and the LC director)
and the associated eigenvector gives the
macroscopic director $\hat{\vect{n}}$. 
A value of $S_2=0$ corresponds to a non-oriented isotropic phase, while
$S_2=1$ to a perfectly aligned mesophase.

\begin{figure}
	\centering
	\includegraphics[width=1\textwidth]{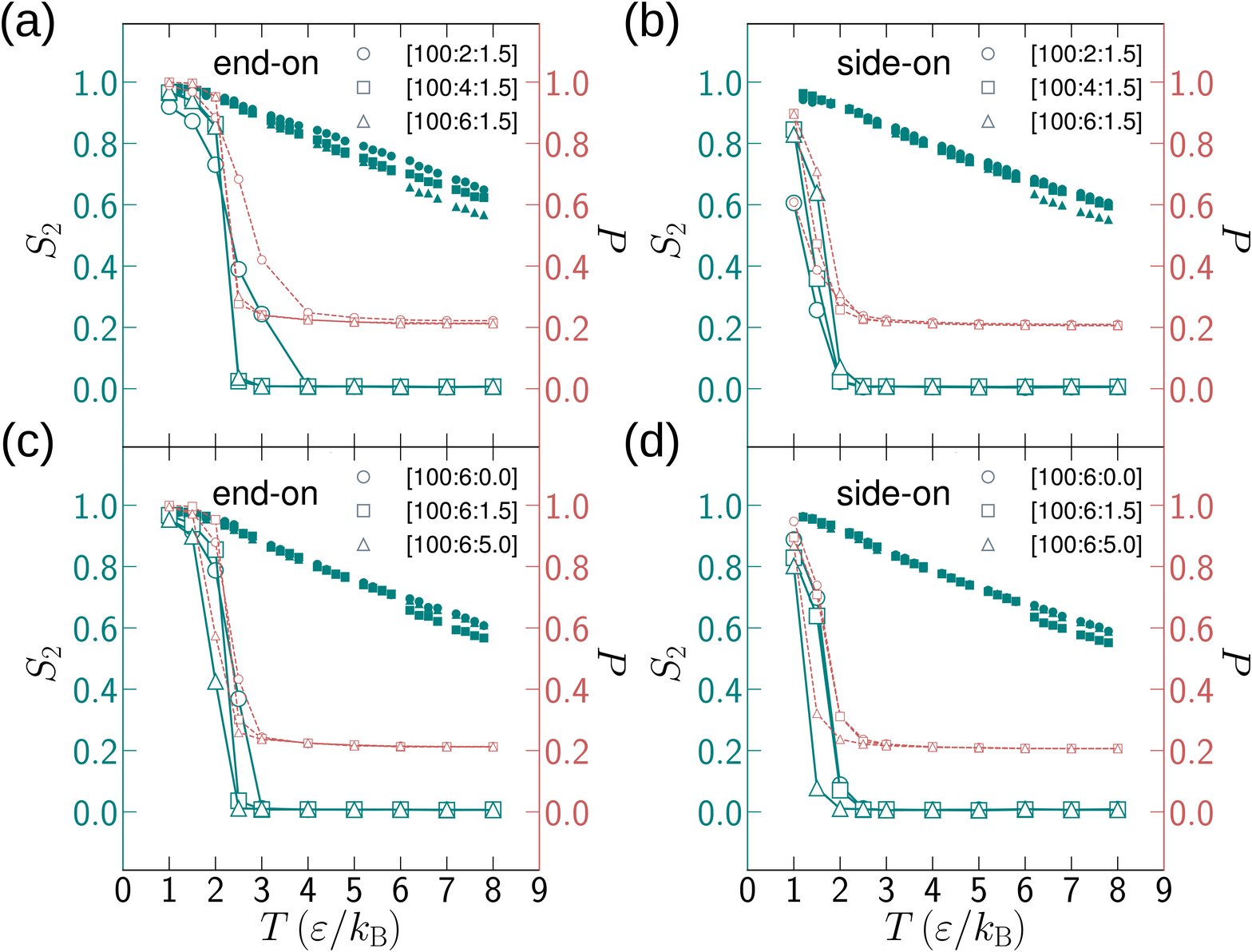}
	\caption{Global orientational order parameter $S_{2}$ (turquoise and continuous lines) 
	for different systems as labeled at different temperatures from 
	equilibrium simulations (empty symbols) after the alignment field 
	is switched off 
	and from cooling runs (filled symbols)
	when the alignment field is switched on.
	The effect of architecture parameters on $S_2$ 
        as a function of temperature
	is shown for different spacer lengths in
        a) for end-on side-chain (EOSC) configurations and in
        b) for side-on side-chain (SOSC) configurations, and
	for different angle force constants ($k_{\beta}$) in
        c) for end-on side-chain (EOSC) configurations and in
        d) for side-on side-chain (SOSC) configurations. 
	Error bars are smaller than the symbols.
	The probability of local orientational order is also shown (red and dashed lines)
	at different temperatures.
	}
	\label{fig:orientation}
\end{figure}

Figure~\ref{fig:orientation} shows the global order parameter 
$S_2$ for EOSC and SOSC systems as labeled, where different variations
of architectural parameters are presented. 
$S_2$ values are shown for relaxation simulations at specific temperatures 
(where the external alignment field is switched off) and for 
cooling simulations (where the external alignment field is switched on)
throughout the process (see Subsec.~\ref{subsec:methodology} for more details).
At high temperatures ($T>3\,\varepsilon/k_{\rm B}$),
the global order parameter, $S_2$, linearly increases as the temperature 
decreases during the cooling runs when the external alignment field 
is switched on. 
Here, $S_2$ values are similar for all cases.
However, when the system is relaxed at a specific temperature (after 
the alignment field is switched off), $S_2$ 
decays to zero as the system returns to its global isotropic order. 
This can be observed in all the different cases studied.
On the other hand, at lower temperatures, $S_2$ starts to increase 
and crystalline order is reached.
For all the cases, the liquid crystal to isotropic transition temperatures 
($T_{\rm LC-I}$) of EOSC systems are consistently higher than those of SOSC systems.
We believe that the reason behind this trend is that
configurational assembly is more favorable for end-on side-chain systems than those for 
side-on side-chain systems 
when the different LC moieties are identically modeled by the same type of ellipsoidal bead
since for end-on LC configuration, the mesogenic unit acts more like 
an extension of the spacer they are attached to.
In experiments \cite{xu2021random},  
the values of the $T_{\rm LC-I}$ may vary depending on the liquid crystal chemistry employed, 
being also possible higher temperatures for side-on side-chain systems.
The reproducibility of specific transition temperatures is outside the scope of this work, 
but we believe that it can be implemented by varying the model parameters.

Some specific architectural trends that can be observed in global ordering are:
(i) SCLCP systems with shorter backbone lengths display 
higher $T_{\rm LC-I}$, or similarly the parameter $S_2$ increases 
its value earlier when the system is cooled down. 
This is valid for both end-on and side-on configurations, 
probably because of fewer steric impediments.
(ii) For systems with different spacer lengths (Figures~\ref{fig:orientation}a,b), 
we observe that the SCLCP system with the shortest side chain 
crystallizes much earlier for the EOSC configuration.
The opposite is observed for the SOSC configuration. 
These trends are maintained as the number of spacer beads increases. 
We infer that in the particular case of systems with spacers composed of 2 beads, 
they do not encounter as many steric hindrances as long side chains, 
but at the same time, the polymer backbone and the mesogen are very coupled,
and for the side-on configuration, the intramolecular interaction 
between the polymer and the LC could be unfavorable for this particular case. 
Furthermore, for both [100:2:1.5] cases, the system is highly elongated, 
which can be generated by this unfavorable tradeoff.
Finally,
(iii) regarding different polymer stiffnesses (Figures~\ref{fig:orientation}c,d), 
the most flexible polymer ($k_{\beta}=0.0\,k_{\rm B}T$) 
is the one with the highest $T_{\rm LC-I}$ for both end-on 
and side-on configurations, 
this is due to the low coupling between the polymer backbone 
and the LC moieties so that the latter can interact 
with the rest of the LCs in the system without major restrictions. 
On the contrary, stiffer systems show 
lower transition temperatures.

Since LC clustering can appear even when global ordering is $\sim 0$,
we quantify local ordering by defining $P$ as the probability of finding
neighbor mesogens with $|\hat{\vect{u}}_{i}\cdot\hat{\vect{u}}_{j}|>0.8$
where $\hat{\vect{u}}_{i}$ and $\hat{\vect{u}}_{j}$ are the unit vectors of 
the main axis of two LCs within a distance less than $3.4\,\sigma$.
$P$ follows the same $S_{2}$ trend for both end-on and side-on configurations 
but shows a higher value, especially at higher temperatures where local 
but not long-range order is present. 

\begin{figure}
	\centering
	\includegraphics[width=1\textwidth]{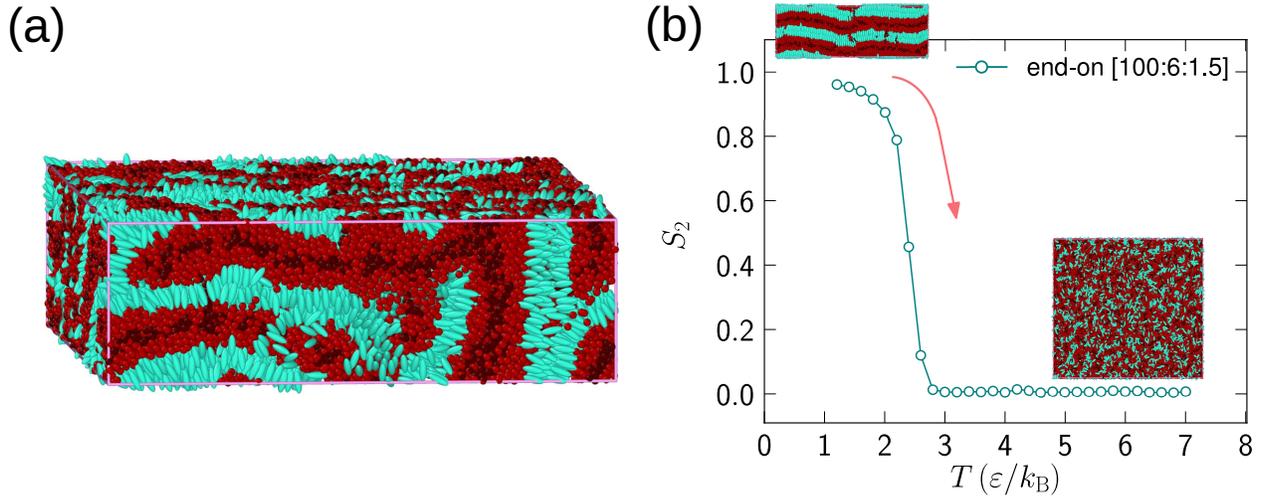}
	\caption{(a) End-on SCLCP [100:6:1.5] configuration 
	at $T=2.0\,\varepsilon/k_{\text B}$ and 
	$P=1.0\,\varepsilon/{\sigma}^{3}$ resulting 
	from a cooling run from isotropic state 
	($T=8.0\,\varepsilon/k_{\text B}$ and 
	$P=1.0\,\varepsilon/{\sigma}^{3}$) without 
	applying the external alignment field to the LC particles.
        (b) Global orientational order parameter $S_{2}$ for the 
        end-on SCLCP [100:6:1.5] configuration at different temperatures 
        resulting from heating the system from smectic B-like to isotropic mesophase without 
	applying the external alignment field.
        A red arrow is depicted to illustrate the direction of the process.
	}
	\label{fig:nofield}
\end{figure}

We have also performed two separate tests for 
the EOSC [100:6:1.5] system.
The first test consisted of cooling this particular system 
from its high-temperature thermodynamic state 
with isotropic structure, where
no external alignment field is applied and the same cooling rate is 
employed in the NPT ensemble. 
The structure at 
$T=2.0\,\varepsilon/k_{\text B}$ and 
$P=1.0\,\varepsilon/{\sigma}^{3}$
is shown in Figure~\ref{fig:nofield}a. 
Here, polysmectic domains are formed, 
with highly local but not global ordering.
In particular, at $T=2\,\varepsilon/k_{\rm B}$, the calculated 
$S_2$ is $\approx 0.16$ and the calculated P is $\approx 0.84$.
In \citet{stimson2005molecular} it is said that without 
an external alignment field the system will evolve 
in the same globally ordered structure 
but longer timescales are needed, which are 
unreachable for MD simulations.
Based on our results, we think that without an
external alignment field there is no specific 
director vector induced and the system will form layers 
that over time will form constraints 
at low temperatures in different directions 
creating entropically trapped systems that will remain 
in a polysmectic mesophase.

In the second test, we started from the 
configuration at 
$T=1.0\,\varepsilon/k_{\text B}$ and 
$P=1.0\,\varepsilon/{\sigma}^{3}$,
which reached that state after applying the procedure explained in
Subsec.~\ref{subsec:methodology},
and we bring it to a thermodynamic state at 
$T=7.0\,\varepsilon/k_{\text B}$ and 
$P=1.0\,\varepsilon/{\sigma}^{3}$ 
by increasing the temperature at a rate of 
$5\times 10^{-5}\,(\varepsilon/k_{\rm B})/\tau$ 
without applying the external alignment field at any point 
of the simulation.
In Figure~\ref{fig:nofield}b it is shown that 
the system returns to a completely isotropic state at high temperatures, 
and that the $S_2$ curve displayed is very similar to that of 
Figure~\ref{fig:orientation}c.

\subsubsection{Radial distribution function}

We have also calculated 
the radial distribution function  
for the liquid crystals, $g_{\rm LC-LC}(r)$,
at different temperatures for all systems studied 
in alignment field-free equilibrium state.
Specifically, $g_{\rm LC-LC}(r)$ is calculated in the standard way,
as the average number of times a LC unit would observe another
LC a distance between $r$ and $r+\Delta r$ from itself, 
normalized by the expected number at uniform density 
(i.e., the number of other LCs in the system divided 
by the system volume). 
The range of $g_{\rm LC-LC}(r)$ is from 0 to large positive numbers. 
It equals 1 indicates random distribution, a  larger value for enrichment, 
and less than 1 for depletion of mesogens at a given separation range.

\begin{figure}
	\centering
	\includegraphics[width=1\textwidth]{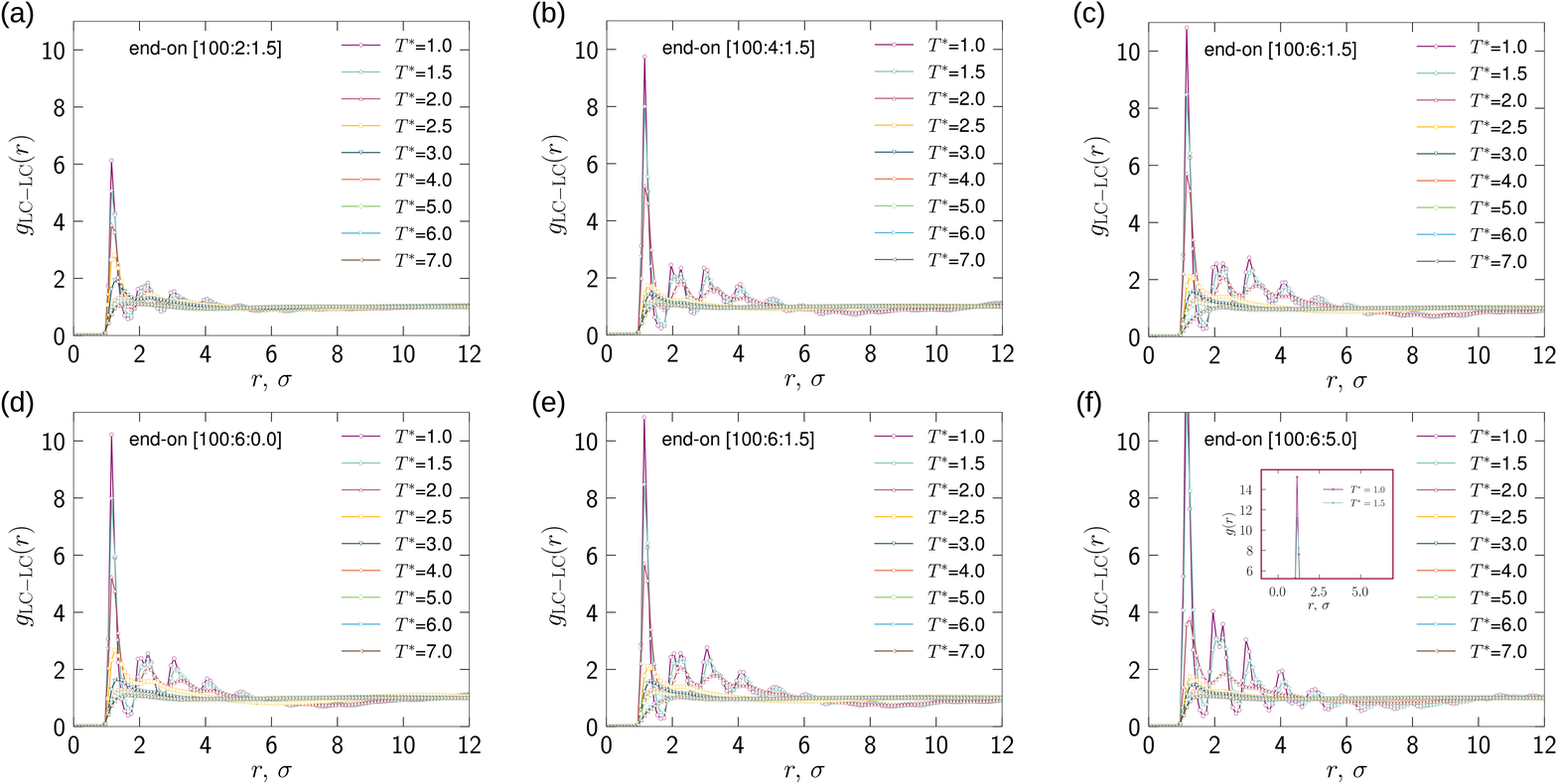}
	\caption{Liquid crystal radial distribution function $g_{\rm LC-LC}(r)$ 
	for all the systems studied at different temperatures from 
	equilibrium simulations.
	Each subplot shows the values only for
	end-on side-chain (EOSC) configuration.
	The effect of architecture parameters on $g_{\rm LC-LC}(r)$ 
	as a function of temperature
	is shown 
	in (a, b, c) for different spacer lengths, and 
	in (d, e, f) for different angle force constants ($k_{\beta}$).  
	}
	\label{fig:gofr_lclc_o}
\end{figure}

In Figure~\ref{fig:gofr_lclc_o}, we show 
the $g_{\rm LC-LC}(r)$ for end-on configuration systems 
at different temperatures and for different architectural parameters.
At high temperatures, the curves look the same, without peaks due 
to the isotropic regime of the systems in this thermodynamic state.
Overall, if we focus on the first (nearest neighbor) peak for every system, 
we can see that this appears at $\sim 2.25\,\sigma$ for temperatures 
below $2.5\,\varepsilon/k_{\rm B}$ for all the systems.
At $T=1.0\,\varepsilon/k_{\rm B}$ we find the highest peak corresponding to the greatest LC packing. 
The second peak is split in the middle which is characteristic of hexagonal lattice arrangement 
as is shown in the top view 
of Figure~\ref{fig:configurations_box}a at low temperature.
Finally, in most cases at low temperatures, 
an undulation extending along the computational box can be observed, 
this reflects and confirms the lamellar smectic B-like order 
in the end-on side-chain systems.

Particularly, if the cases for different backbone lengths 
are compared, 
it can be observed that 
they are very similar to each other in structural terms. 
Now, if we compare cases with different spacer lengths 
(Figure~\ref{fig:gofr_lclc_o}a,b,c), it can be observed that 
the first peak increases as this architectural parameter increases, 
this implies that the greater the LC-backbone decoupling, 
the greater the ordering and packing of the mesogens. 
It can also be observed that the particular case of EOSC [100:2:1.5] 
(Figure~\ref{fig:gofr_lclc_o}a), 
presents neither the division of the second peak split corresponding to hexagonal order 
nor undulation at long distances, which is due to the high elongation 
of the system and its high $T_{\rm LC-I}$ which could cause 
probable artifacts in the results.
For different polymer stiffnesses (Figure~\ref{fig:gofr_lclc_o}d,e,f), 
the higher the polymer stiffness, the higher the first peak 
at low temperatures but at the same time a noisier long-distance 
undulation because the smectic B-like order formed at low temperatures 
for EOSC [100:6: 5.0] (Figure~\ref{fig:gofr_lclc_o}f), 
is only composed of a monolayer and not of a bilayer of LC moieties.

\begin{figure}
	\centering
	\includegraphics[width=1\textwidth]{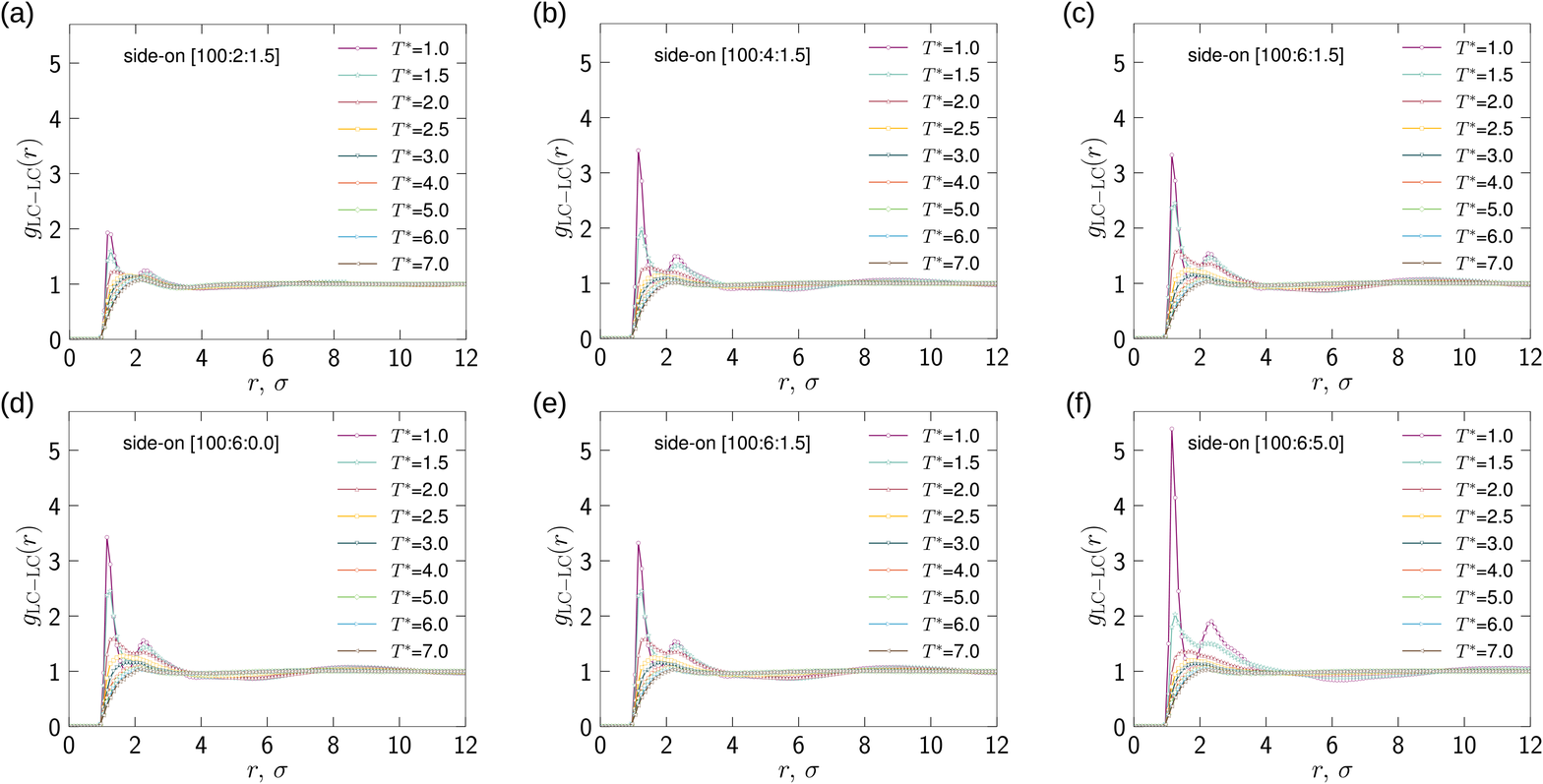}
	\caption{Liquid crystal radial distribution function $g_{\rm LC-LC}(r)$ 
	for all the systems studied at different temperatures from 
	equilibrium simulations.
	Each subplot shows the values only for
	side-on side-chain (SOSC) configuration.
	The effect of architecture parameters on $g_{\rm LC-LC}(r)$ 
	as a function of temperature
	is shown 
	in (a, b, c) for different spacer lengths, and 
	in (d, e, f) for different angle force constants ($k_{\beta}$).  
	}
	\label{fig:gofr_lclc_p}
\end{figure}

All plots for the side-on SCLCP systems have similar shapes as shown 
in Figure~\ref{fig:gofr_lclc_p}, with peaks and local maxima 
forming at temperatures lower than the crystalline mesophase transition temperature.
If compared with EOSC systems, the first and second peaks 
are much smaller for SOSC cases due to the absence of smectic order.
Also, patterns indicative of a hexagonal lattice arrangement along planes orthogonal 
to the layering direction exhibited in EOSC systems 
(second peak splits in the $g_{\rm LC-LC}$) 
are no longer displayed here.
The undulations showed at low temperatures for long distances 
are a result of the hexagonal-like shape caused by the LC moieties  
surrounding and isolating the polymer chains.
In both the end-on and side-on configurations, the [100:6:5.0] case
demonstrated the most prominent peaks and the [100:2:1.5] case
exhibited the least prominent peaks, the latter being probably 
resultant of artifacts in the measurements due to 
the high elongation of the computational box.

\section{PERFORMANCE}
\label{sec:performance}

We have measured the computational performance 
of the coarse-grained SCLCP model.
To do this, we have made different comparisons in 
terms of architectural parameters.
In Table~\ref{tab:performance}, the fist comparison is between two melts 
of linear polymer chains modeled in different ways.
The first system modeled with the worm-like chain model, 
like the rest of the systems in this work, is $\sim 24\%$ 
less efficient than a melt composed of the same number of linear 
polymeric chains but modeled using the Kremer--Grest (KG) model 
\cite{kremer1990dynamics}. 
This result confirms what was expected, 
a model with more degrees of freedom will be more expensive 
in computational terms, but $\sim 24\%$ is feasible
taking into account the advantages of greater control 
over the system.
Note that a time step of $0.002\,\tau$ was used in both simulations, 
although generally the time step for the KG model where the beads 
are connected along the chain via the finitely extensible
nonlinear elastic (FENE), can be safely increased to at least $0.005\,\tau$, 
and although for the model used in this work larger timesteps can be used, 
it would hardly reach $0.005\,\tau$ safely, 
therefore the difference in efficiency would be greater.
The second comparison in Table~\ref{tab:performance} basically shows how 
much computational efficiency is lost by using the orientational potential 
to implement syndiotacticity as it is used in simulations of this work. 
The result shows only a $\sim 1.4\%$ difference 
when compared with the same model but atactic, 
that is, the intramolecular interactions that set 
this are deactivated (see Subsec.~\ref{subsec:model} for details
on this intramolecular interaction). 
Therefore, including tacticity in the model 
does not incur in a considerable computational expense.
The last comparison in Table~\ref{tab:performance} 
in terms of computational performance 
was made for a melt composed of linear chains [100:0:1.5] 
and a system of SCLCPs [100:6:1.5] both with the same number of molecules, 
however, in the latter, there are more particles
(112 cores in parallel are employed for this comparison). 
The linear polymeric chains of both systems are modeled 
with the same worm-like chain model, 
but the SCLCP system includes the syndiotactically attached 
spacers and the mesogenic units.
Here, the difference in computational efficiency is about a $\sim 915\%$. 
This huge difference is due to the inclusion of  
mesogenic units modeled as uniaxial ellipsoids through 
the Gay--Berne potential, 
which is a very expensive potential and responsible 
for the decrease in computational performance, 
although one of the best options available to model 
anisotropic interactions.
Finally, we have explored the computational scalability, 
performing simulations of SCLCP systems (for EOSC and SOSC configurations) 
with different numbers of cores, 
which is close to ideal.


\begin{table}
\caption{Comparison of the computational performance of different features 
of the SCLCP model.
All the measurements are performed at $T=8.0\,\varepsilon/k_{\rm B}$, 
$P=1.0\,\varepsilon/{\sigma}^{3}$, and a timestep of $0.002\,\tau$.
The percentage difference is defined as how much efficient is the first system listed in comparison
with the second system listed for each set of different comparisons.}
\label{tab:performance}
\begin{tabular}{llc}
\hline
\hline
 &  &  {\textbf{Computational}}\\ 
 &  &  {\textbf{performance}}\\ 
\hline
{\textbf{System}} & linear (Kremer--Grest) [100:0:1.5]      & $262135\,\tau$/day \\ 
                  & linear [100:0:1.5] & $211329\,\tau$/day \\ 
\hline
Percentage difference:  24\%  & Number of cores:  56 & \\  
\hline
\hline
 &  &  {\textbf{Computational}}\\ 
 &  &  {\textbf{performance}}\\ 
\hline
{\textbf{System}} & end-on  [100:6:1.5] (atactic)      & $15599\,\tau$/day \\ 
                  & end-on  [100:6:1.5] (syndiotactic)  & $15380\,\tau$/day \\ 
\hline
Percentage difference:  1.4\%   & Number of cores: 56  &\\  
\hline
\hline
 &  &  {\textbf{Computational}}\\ 
 &  &  {\textbf{performance}}\\ 
\hline
{\textbf{System}} & linear [100:0:1.5]             & $328362\,\tau$/day \\ 
                  & end-on  [100:6:1.5]  & $32356\,\tau$/day \\ 
\hline
Percentage difference:   915\%  & Number of cores: 112 &  \\  
\hline
\hline
\end{tabular}
\end{table}

\section{CONCLUSION}

In this work, we have developed a coarse-grained model 
for side-chain liquid crystal polymers.
The model is well-defined in orientational terms, 
which allows constraints to be set to avoid unphysical rotation behavior 
of the mesogenic groups attached to the spacer chains, 
reproduce specific tacticity, and 
make the model responsive to external stimuli,
such as external alignment fields.
The model can be easily modified 
to implement different types of liquid
crystalline sequences and compositions.
The architectural flexibility of the model arises as a consequence 
of the increased degrees of freedom of its constituents, 
however, it is still computationally efficient enough 
to simulate large size and long time scales and, in particular, 
to study different LC mesophases.
The model was tested with extensive simulations
using standard parameters for nonbonded interactions 
and potentials in dimensionless units, 
and also standard parametrizations 
for ellipsoidal particles that reproduce the anisotropic behavior of mesogens.
Although, the model has the potential to map a specific chemistry.
This is especially important since, as we are aware, 
this is the first coarse-grained model that can qualitatively reproduce 
the transition temperature difference 
between purely side-on and purely end-on configurations;
by mapping to specific chemistry, we can move towards 
quantitative predictions of this transition.
Finally, we can anticipate that the model will be useful
to study a series of sequences and compositions of LC groups along 
the chain that can allow for a material with
multiple types of deformations as a function of temperature.
Different architectures, such as cross-linked bottlebrushes configurations 
will be explored in future work.
These outcomes could serve as inspiration for 
applications with soft responsive materials.

\begin{acknowledgement}

Acknowledgment is made to the Donors of 
the American Chemical Society Petroleum Research Fund for support 
of this research.
We also thank the Ohio Supercomputer Center
for high-performance computing resources.
(\url{http://osc.edu/ark:/19495/f5s1ph73})
\cite{OhioSupercomputerCenter1987}.

\end{acknowledgement}

\begin{suppinfo}

The data that support the findings of this study are available 
within the article and its supplementary material.

The external alignment field implemented as a fix \cite{becerra2022fixext} for the LAMMPS package and 
its documentation \href{https://dbecerra-dev.github.io/fix-ext_alignment}{\tt https://dbecerra-dev.github.io/fix-ext\_alignment} 
are publicly available.

\end{suppinfo}

\bibliography{aipsamp}

\providecommand{\latin}[1]{#1}
\makeatletter
\providecommand{\doi}
  {\begingroup\let\do\@makeother\dospecials
  \catcode`\{=1 \catcode`\}=2 \doi@aux}
\providecommand{\doi@aux}[1]{\endgroup\texttt{#1}}
\makeatother
\providecommand*\mcitethebibliography{\thebibliography}
\csname @ifundefined\endcsname{endmcitethebibliography}
  {\let\endmcitethebibliography\endthebibliography}{}
\begin{mcitethebibliography}{82}
\providecommand*\natexlab[1]{#1}
\providecommand*\mciteSetBstSublistMode[1]{}
\providecommand*\mciteSetBstMaxWidthForm[2]{}
\providecommand*\mciteBstWouldAddEndPuncttrue
  {\def\EndOfBibitem{\unskip.}}
\providecommand*\mciteBstWouldAddEndPunctfalse
  {\let\EndOfBibitem\relax}
\providecommand*\mciteSetBstMidEndSepPunct[3]{}
\providecommand*\mciteSetBstSublistLabelBeginEnd[3]{}
\providecommand*\EndOfBibitem{}
\mciteSetBstSublistMode{f}
\mciteSetBstMaxWidthForm{subitem}{(\alph{mcitesubitemcount})}
\mciteSetBstSublistLabelBeginEnd
  {\mcitemaxwidthsubitemform\space}
  {\relax}
  {\relax}

\bibitem[Allen and Wilson(1989)Allen, and Wilson]{allen1989computer}
Allen,~M.~P.; Wilson,~M.~R. Computer simulation of liquid crystals.
  \emph{Journal of Computer-Aided Molecular Design} \textbf{1989}, \emph{3},
  335--353\relax
\mciteBstWouldAddEndPuncttrue
\mciteSetBstMidEndSepPunct{\mcitedefaultmidpunct}
{\mcitedefaultendpunct}{\mcitedefaultseppunct}\relax
\EndOfBibitem
\bibitem[Lu \latin{et~al.}(2017)Lu, Guo, Tong, Xia, and Zhao]{lu2017tunable}
Lu,~X.; Guo,~S.; Tong,~X.; Xia,~H.; Zhao,~Y. Tunable photocontrolled motions
  using stored strain energy in malleable azobenzene liquid crystalline polymer
  actuators. \emph{Advanced Materials} \textbf{2017}, \emph{29}, 1606467\relax
\mciteBstWouldAddEndPuncttrue
\mciteSetBstMidEndSepPunct{\mcitedefaultmidpunct}
{\mcitedefaultendpunct}{\mcitedefaultseppunct}\relax
\EndOfBibitem
\bibitem[Yang and Zhao(2018)Yang, and Zhao]{yang2018multitemperature}
Yang,~R.; Zhao,~Y. Multitemperature memory actuation of a liquid crystal
  polymer network over a broad nematic--isotropic phase transition induced by
  large Strain. \emph{ACS Macro Letters} \textbf{2018}, \emph{7},
  353--357\relax
\mciteBstWouldAddEndPuncttrue
\mciteSetBstMidEndSepPunct{\mcitedefaultmidpunct}
{\mcitedefaultendpunct}{\mcitedefaultseppunct}\relax
\EndOfBibitem
\bibitem[Bellin \latin{et~al.}(2006)Bellin, Kelch, Langer, and
  Lendlein]{bellin2006polymeric}
Bellin,~I.; Kelch,~S.; Langer,~R.; Lendlein,~A. Polymeric triple-shape
  materials. \emph{Proceedings of the National Academy of Sciences}
  \textbf{2006}, \emph{103}, 18043--18047\relax
\mciteBstWouldAddEndPuncttrue
\mciteSetBstMidEndSepPunct{\mcitedefaultmidpunct}
{\mcitedefaultendpunct}{\mcitedefaultseppunct}\relax
\EndOfBibitem
\bibitem[Ahir \latin{et~al.}(2006)Ahir, Tajbakhsh, and Terentjev]{ahir2006self}
Ahir,~S.~V.; Tajbakhsh,~A.~R.; Terentjev,~E.~M. Self-assembled shape-memory
  fibers of triblock liquid-crystal polymers. \emph{Advanced Functional
  Materials} \textbf{2006}, \emph{16}, 556--560\relax
\mciteBstWouldAddEndPuncttrue
\mciteSetBstMidEndSepPunct{\mcitedefaultmidpunct}
{\mcitedefaultendpunct}{\mcitedefaultseppunct}\relax
\EndOfBibitem
\bibitem[Qin and Mather(2009)Qin, and Mather]{qin2009combined}
Qin,~H.; Mather,~P.~T. Combined one-way and two-way shape memory in a
  glass-forming nematic network. \emph{Macromolecules} \textbf{2009},
  \emph{42}, 273--280\relax
\mciteBstWouldAddEndPuncttrue
\mciteSetBstMidEndSepPunct{\mcitedefaultmidpunct}
{\mcitedefaultendpunct}{\mcitedefaultseppunct}\relax
\EndOfBibitem
\bibitem[Ware \latin{et~al.}(2016)Ware, Biggins, Shick, Warner, and
  White]{ware2016localized}
Ware,~T.~H.; Biggins,~J.~S.; Shick,~A.~F.; Warner,~M.; White,~T.~J. Localized
  soft elasticity in liquid crystal elastomers. \emph{Nature communications}
  \textbf{2016}, \emph{7}, 1--7\relax
\mciteBstWouldAddEndPuncttrue
\mciteSetBstMidEndSepPunct{\mcitedefaultmidpunct}
{\mcitedefaultendpunct}{\mcitedefaultseppunct}\relax
\EndOfBibitem
\bibitem[Xu \latin{et~al.}(2021)Xu, Dupont, Yao, Zhang, Fang, and
  Wang]{xu2021random}
Xu,~Y.; Dupont,~R.~L.; Yao,~Y.; Zhang,~M.; Fang,~J.-C.; Wang,~X. Random Liquid
  Crystalline Copolymers Consisting of Prolate and Oblate Liquid Crystal
  Monomers. \emph{Macromolecules} \textbf{2021}, \emph{54}, 5376--5387\relax
\mciteBstWouldAddEndPuncttrue
\mciteSetBstMidEndSepPunct{\mcitedefaultmidpunct}
{\mcitedefaultendpunct}{\mcitedefaultseppunct}\relax
\EndOfBibitem
\bibitem[De~Gennes(1969)]{de1969possibilites}
De~Gennes,~P. Possibilites offertes par la reticulation de polymeres en
  presence d'un cristal liquide. \emph{Physics Letters A} \textbf{1969},
  \emph{28}, 725--726\relax
\mciteBstWouldAddEndPuncttrue
\mciteSetBstMidEndSepPunct{\mcitedefaultmidpunct}
{\mcitedefaultendpunct}{\mcitedefaultseppunct}\relax
\EndOfBibitem
\bibitem[Warner and Terentjev(2007)Warner, and Terentjev]{warner2007liquid}
Warner,~M.; Terentjev,~E.~M. \emph{Liquid crystal elastomers}; Oxford
  university press, 2007; Vol. 120\relax
\mciteBstWouldAddEndPuncttrue
\mciteSetBstMidEndSepPunct{\mcitedefaultmidpunct}
{\mcitedefaultendpunct}{\mcitedefaultseppunct}\relax
\EndOfBibitem
\bibitem[Behl and Lendlein(2007)Behl, and Lendlein]{behl2007shape}
Behl,~M.; Lendlein,~A. Shape-memory polymers. \emph{Materials Today}
  \textbf{2007}, \emph{10}, 20--28\relax
\mciteBstWouldAddEndPuncttrue
\mciteSetBstMidEndSepPunct{\mcitedefaultmidpunct}
{\mcitedefaultendpunct}{\mcitedefaultseppunct}\relax
\EndOfBibitem
\bibitem[Ahn \latin{et~al.}(2011)Ahn, Deshmukh, Gopinadhan, Osuji, and
  Kasi]{ahn2011side}
Ahn,~S.; Deshmukh,~P.; Gopinadhan,~M.; Osuji,~C.~O.; Kasi,~R.~M. Side-chain
  liquid crystalline polymer networks: exploiting nanoscale smectic
  polymorphism to design shape-memory polymers. \emph{ACS Nano} \textbf{2011},
  \emph{5}, 3085--3095\relax
\mciteBstWouldAddEndPuncttrue
\mciteSetBstMidEndSepPunct{\mcitedefaultmidpunct}
{\mcitedefaultendpunct}{\mcitedefaultseppunct}\relax
\EndOfBibitem
\bibitem[Shenoy \latin{et~al.}(2002)Shenoy, Thomsen~III, Srinivasan, Keller,
  and Ratna]{shenoy2002carbon}
Shenoy,~D.~K.; Thomsen~III,~D.~L.; Srinivasan,~A.; Keller,~P.; Ratna,~B.~R.
  Carbon coated liquid crystal elastomer film for artificial muscle
  applications. \emph{Sensors and Actuators A: Physical} \textbf{2002},
  \emph{96}, 184--188\relax
\mciteBstWouldAddEndPuncttrue
\mciteSetBstMidEndSepPunct{\mcitedefaultmidpunct}
{\mcitedefaultendpunct}{\mcitedefaultseppunct}\relax
\EndOfBibitem
\bibitem[Spillmann \latin{et~al.}(2007)Spillmann, Naciri, Martin, Farahat,
  Herr, and Ratna]{spillmann2007stacking}
Spillmann,~C.~M.; Naciri,~J.; Martin,~B.~D.; Farahat,~W.; Herr,~H.;
  Ratna,~B.~R. Stacking nematic elastomers for artificial muscle applications.
  \emph{Sensors and Actuators A: Physical} \textbf{2007}, \emph{133},
  500--505\relax
\mciteBstWouldAddEndPuncttrue
\mciteSetBstMidEndSepPunct{\mcitedefaultmidpunct}
{\mcitedefaultendpunct}{\mcitedefaultseppunct}\relax
\EndOfBibitem
\bibitem[Li and Keller(2006)Li, and Keller]{li2006artificial}
Li,~M.~H.; Keller,~P. Artificial muscles based on liquid crystal elastomers.
  \emph{Philosophical Transactions of the Royal Society A: Mathematical,
  Physical and Engineering Sciences} \textbf{2006}, \emph{364},
  2763--2777\relax
\mciteBstWouldAddEndPuncttrue
\mciteSetBstMidEndSepPunct{\mcitedefaultmidpunct}
{\mcitedefaultendpunct}{\mcitedefaultseppunct}\relax
\EndOfBibitem
\bibitem[Kim \latin{et~al.}(2019)Kim, Park, Won, Jeon, and
  Wie]{kim2019contactless}
Kim,~J.~G.; Park,~J.~E.; Won,~S.; Jeon,~J.; Wie,~J.~J. Contactless manipulation
  of soft robots. \emph{Materials} \textbf{2019}, \emph{12}, 3065\relax
\mciteBstWouldAddEndPuncttrue
\mciteSetBstMidEndSepPunct{\mcitedefaultmidpunct}
{\mcitedefaultendpunct}{\mcitedefaultseppunct}\relax
\EndOfBibitem
\bibitem[Wang \latin{et~al.}(2022)Wang, Liu, and Yang]{wang2022multi}
Wang,~Y.; Liu,~J.; Yang,~S. Multi-functional liquid crystal elastomer
  composites. \emph{Applied Physics Reviews} \textbf{2022}, \emph{9},
  011301\relax
\mciteBstWouldAddEndPuncttrue
\mciteSetBstMidEndSepPunct{\mcitedefaultmidpunct}
{\mcitedefaultendpunct}{\mcitedefaultseppunct}\relax
\EndOfBibitem
\bibitem[Kempe \latin{et~al.}(2004)Kempe, Scruggs, Verduzco, Lal, and
  Kornfield]{kempe2004self}
Kempe,~M.~D.; Scruggs,~N.~R.; Verduzco,~R.; Lal,~J.; Kornfield,~J.~A.
  Self-assembled liquid-crystalline gels designed from the bottom up.
  \emph{Nature Materials} \textbf{2004}, \emph{3}, 177--182\relax
\mciteBstWouldAddEndPuncttrue
\mciteSetBstMidEndSepPunct{\mcitedefaultmidpunct}
{\mcitedefaultendpunct}{\mcitedefaultseppunct}\relax
\EndOfBibitem
\bibitem[Kempe \latin{et~al.}(2006)Kempe, Verduzco, Scruggs, and
  Kornfield]{kempe2006rheological}
Kempe,~M.~D.; Verduzco,~R.; Scruggs,~N.~R.; Kornfield,~J.~A. Rheological study
  of structural transitions in triblock copolymers in a liquid crystal solvent.
  \emph{Soft Matter} \textbf{2006}, \emph{2}, 422--431\relax
\mciteBstWouldAddEndPuncttrue
\mciteSetBstMidEndSepPunct{\mcitedefaultmidpunct}
{\mcitedefaultendpunct}{\mcitedefaultseppunct}\relax
\EndOfBibitem
\bibitem[Thakur \latin{et~al.}(2016)Thakur, Kessler, \latin{et~al.}
  others]{thakur2016liquid}
Thakur,~V.~K.; Kessler,~M.~R., \latin{et~al.}  \emph{Liquid crystalline
  polymers}; Springer, 2016\relax
\mciteBstWouldAddEndPuncttrue
\mciteSetBstMidEndSepPunct{\mcitedefaultmidpunct}
{\mcitedefaultendpunct}{\mcitedefaultseppunct}\relax
\EndOfBibitem
\bibitem[Moore and Stupp(1987)Moore, and Stupp]{moore1987orientation}
Moore,~J.; Stupp,~S. Orientation dynamics of main-chain liquid crystal
  polymers. 2. Structure and kinetics in a magnetic field.
  \emph{Macromolecules} \textbf{1987}, \emph{20}, 282--293\relax
\mciteBstWouldAddEndPuncttrue
\mciteSetBstMidEndSepPunct{\mcitedefaultmidpunct}
{\mcitedefaultendpunct}{\mcitedefaultseppunct}\relax
\EndOfBibitem
\bibitem[Affouard \latin{et~al.}(1996)Affouard, Kr{\"o}ger, and
  Hess]{affouard1996molecular}
Affouard,~F.; Kr{\"o}ger,~M.; Hess,~S. Molecular dynamics of model liquid
  crystals composed of semiflexible molecules. \emph{Physical Review E}
  \textbf{1996}, \emph{54}, 5178\relax
\mciteBstWouldAddEndPuncttrue
\mciteSetBstMidEndSepPunct{\mcitedefaultmidpunct}
{\mcitedefaultendpunct}{\mcitedefaultseppunct}\relax
\EndOfBibitem
\bibitem[Lyulin \latin{et~al.}(1998)Lyulin, Al-Barwani, Allen, Wilson, Neelov,
  and Allsopp]{lyulin1998molecular}
Lyulin,~A.~V.; Al-Barwani,~M.~S.; Allen,~M.~P.; Wilson,~M.~R.; Neelov,~I.;
  Allsopp,~N.~K. Molecular dynamics simulation of main chain liquid crystalline
  polymers. \emph{Macromolecules} \textbf{1998}, \emph{31}, 4626--4634\relax
\mciteBstWouldAddEndPuncttrue
\mciteSetBstMidEndSepPunct{\mcitedefaultmidpunct}
{\mcitedefaultendpunct}{\mcitedefaultseppunct}\relax
\EndOfBibitem
\bibitem[Chen \latin{et~al.}(2016)Chen, Zhu, Cui, Liu, Sun, Chen, and
  Li]{chen2016gpu}
Chen,~W.; Zhu,~Y.; Cui,~F.; Liu,~L.; Sun,~Z.; Chen,~J.; Li,~Y. GPU-accelerated
  molecular dynamics simulation to study liquid crystal phase transition using
  coarse-grained gay-berne anisotropic potential. \emph{PLOS One}
  \textbf{2016}, \emph{11}, e0151704\relax
\mciteBstWouldAddEndPuncttrue
\mciteSetBstMidEndSepPunct{\mcitedefaultmidpunct}
{\mcitedefaultendpunct}{\mcitedefaultseppunct}\relax
\EndOfBibitem
\bibitem[Cuierrier \latin{et~al.}(2021)Cuierrier, Ebrahimi, Couture, and
  Soldera]{cuierrier2021simulation}
Cuierrier,~E.; Ebrahimi,~S.; Couture,~O.; Soldera,~A. Simulation of main chain
  liquid crystalline polymers using a Gay-Berne/Lennard-Jones hybrid model.
  \emph{Computational Materials Science} \textbf{2021}, \emph{186},
  110041\relax
\mciteBstWouldAddEndPuncttrue
\mciteSetBstMidEndSepPunct{\mcitedefaultmidpunct}
{\mcitedefaultendpunct}{\mcitedefaultseppunct}\relax
\EndOfBibitem
\bibitem[Berardi \latin{et~al.}(2004)Berardi, Micheletti, Muccioli, Ricci, and
  Zannoni]{berardi2004computer}
Berardi,~R.; Micheletti,~D.; Muccioli,~L.; Ricci,~M.; Zannoni,~C. A computer
  simulation study of the influence of a liquid crystal medium on
  polymerization. \emph{The Journal of Chemical Physics} \textbf{2004},
  \emph{121}, 9123--9130\relax
\mciteBstWouldAddEndPuncttrue
\mciteSetBstMidEndSepPunct{\mcitedefaultmidpunct}
{\mcitedefaultendpunct}{\mcitedefaultseppunct}\relax
\EndOfBibitem
\bibitem[Stimson and Wilson(2005)Stimson, and Wilson]{stimson2005molecular}
Stimson,~L.~M.; Wilson,~M.~R. Molecular dynamics simulations of side chain
  liquid crystal polymer molecules in isotropic and liquid-crystalline melts.
  \emph{The Journal of Chemical Physics} \textbf{2005}, \emph{123},
  034908\relax
\mciteBstWouldAddEndPuncttrue
\mciteSetBstMidEndSepPunct{\mcitedefaultmidpunct}
{\mcitedefaultendpunct}{\mcitedefaultseppunct}\relax
\EndOfBibitem
\bibitem[Collings and Hird(2017)Collings, and Hird]{collings2017introduction}
Collings,~P.~J.; Hird,~M. \emph{Introduction to liquid crystals chemistry and
  physics}; CRC Press, 2017\relax
\mciteBstWouldAddEndPuncttrue
\mciteSetBstMidEndSepPunct{\mcitedefaultmidpunct}
{\mcitedefaultendpunct}{\mcitedefaultseppunct}\relax
\EndOfBibitem
\bibitem[Kannan \latin{et~al.}(1993)Kannan, Kornfield, Schwenk, and
  Boeffel]{kannan1993rheology}
Kannan,~R.~M.; Kornfield,~J.~A.; Schwenk,~N.; Boeffel,~C. Rheology of
  side-group liquid-crystalline polymers: effect of isotropic-nematic
  transition and evidence of flow alignment. \emph{Macromolecules}
  \textbf{1993}, \emph{26}, 2050--2056\relax
\mciteBstWouldAddEndPuncttrue
\mciteSetBstMidEndSepPunct{\mcitedefaultmidpunct}
{\mcitedefaultendpunct}{\mcitedefaultseppunct}\relax
\EndOfBibitem
\bibitem[d'Allest \latin{et~al.}(1988)d'Allest, Maissa, Ten~Bosch, Sixou,
  Blumstein, Blumstein, Teixeira, and Noirez]{d1988experimental}
d'Allest,~J.; Maissa,~P.; Ten~Bosch,~A.; Sixou,~P.; Blumstein,~A.;
  Blumstein,~R.; Teixeira,~J.; Noirez,~L. Experimental evidence of chain
  extension at the transition temperature of a nematic polymer. \emph{Physical
  Review Letters} \textbf{1988}, \emph{61}, 2562\relax
\mciteBstWouldAddEndPuncttrue
\mciteSetBstMidEndSepPunct{\mcitedefaultmidpunct}
{\mcitedefaultendpunct}{\mcitedefaultseppunct}\relax
\EndOfBibitem
\bibitem[Wewerka \latin{et~al.}(2001)Wewerka, Floudas, Pakula, and
  Stelzer]{wewerka2001side}
Wewerka,~A.; Floudas,~G.; Pakula,~T.; Stelzer,~F. Side-chain liquid-crystalline
  homopolymers and copolymers. Structure and rheology. \emph{Macromolecules}
  \textbf{2001}, \emph{34}, 8129--8137\relax
\mciteBstWouldAddEndPuncttrue
\mciteSetBstMidEndSepPunct{\mcitedefaultmidpunct}
{\mcitedefaultendpunct}{\mcitedefaultseppunct}\relax
\EndOfBibitem
\bibitem[Wewerka \latin{et~al.}(2001)Wewerka, Viertler, Vlassopoulos, and
  Stelzer]{wewerka2001structure}
Wewerka,~A.; Viertler,~K.; Vlassopoulos,~D.; Stelzer,~F. Structure and rheology
  of model side-chain liquid crystalline polymers with varying mesogen length.
  \emph{Rheologica Acta} \textbf{2001}, \emph{40}, 416--425\relax
\mciteBstWouldAddEndPuncttrue
\mciteSetBstMidEndSepPunct{\mcitedefaultmidpunct}
{\mcitedefaultendpunct}{\mcitedefaultseppunct}\relax
\EndOfBibitem
\bibitem[Rendon \latin{et~al.}(2007)Rendon, Burghardt, Auad, and
  Kornfield]{rendon2007shear}
Rendon,~S.; Burghardt,~W.~R.; Auad,~M.~L.; Kornfield,~J.~A. Shear-induced
  alignment of smectic side group liquid crystalline polymers.
  \emph{Macromolecules} \textbf{2007}, \emph{40}, 6624--6630\relax
\mciteBstWouldAddEndPuncttrue
\mciteSetBstMidEndSepPunct{\mcitedefaultmidpunct}
{\mcitedefaultendpunct}{\mcitedefaultseppunct}\relax
\EndOfBibitem
\bibitem[Colby \latin{et~al.}(1993)Colby, Gillmor, Galli, Laus, Ober, and
  Hall]{colby1993linear}
Colby,~R.~H.; Gillmor,~J.; Galli,~G.; Laus,~M.; Ober,~C.; Hall,~E. Linear
  viscoelasticity of side chain liquid crystal polymer. \emph{Liquid Crystals}
  \textbf{1993}, \emph{13}, 233--245\relax
\mciteBstWouldAddEndPuncttrue
\mciteSetBstMidEndSepPunct{\mcitedefaultmidpunct}
{\mcitedefaultendpunct}{\mcitedefaultseppunct}\relax
\EndOfBibitem
\bibitem[Pleiner and Brand(1992)Pleiner, and Brand]{pleiner1992local}
Pleiner,~H.; Brand,~H.~R. Local rotational degrees of freedom in nematic
  liquid-crystalline side-chain polymers. \emph{Macromolecules} \textbf{1992},
  \emph{25}, 895--901\relax
\mciteBstWouldAddEndPuncttrue
\mciteSetBstMidEndSepPunct{\mcitedefaultmidpunct}
{\mcitedefaultendpunct}{\mcitedefaultseppunct}\relax
\EndOfBibitem
\bibitem[Fourmaux-Demange \latin{et~al.}(1998)Fourmaux-Demange, Br{\^u}let,
  Cotton, Hilliou, Martinoty, Keller, and Bou{\'e}]{fourmaux1998rheology}
Fourmaux-Demange,~V.; Br{\^u}let,~A.; Cotton,~J.; Hilliou,~L.; Martinoty,~P.;
  Keller,~P.; Bou{\'e},~F. Rheology of a comblike liquid crystalline polymer as
  a function of its molecular weight. \emph{Macromolecules} \textbf{1998},
  \emph{31}, 7445--7452\relax
\mciteBstWouldAddEndPuncttrue
\mciteSetBstMidEndSepPunct{\mcitedefaultmidpunct}
{\mcitedefaultendpunct}{\mcitedefaultseppunct}\relax
\EndOfBibitem
\bibitem[Long and Morse(2002)Long, and Morse]{long2002rouse}
Long,~D.; Morse,~D.~C. A Rouse-like model of liquid crystalline polymer melts:
  Director dynamics and linear viscoelasticity. \emph{Journal of Rheology}
  \textbf{2002}, \emph{46}, 49--92\relax
\mciteBstWouldAddEndPuncttrue
\mciteSetBstMidEndSepPunct{\mcitedefaultmidpunct}
{\mcitedefaultendpunct}{\mcitedefaultseppunct}\relax
\EndOfBibitem
\bibitem[Wang and Warner(1987)Wang, and Warner]{wang1987theory}
Wang,~X.; Warner,~M. Theory of nematic comb-like polymers. \emph{Journal of
  Physics A: Mathematical and General} \textbf{1987}, \emph{20}, 713\relax
\mciteBstWouldAddEndPuncttrue
\mciteSetBstMidEndSepPunct{\mcitedefaultmidpunct}
{\mcitedefaultendpunct}{\mcitedefaultseppunct}\relax
\EndOfBibitem
\bibitem[Renz and Warner(1988)Renz, and Warner]{renz1988theory}
Renz,~W.; Warner,~M. The theory of competing nematic phases of comb polymers.
  \emph{Proceedings of the Royal Society of London. A. Mathematical and
  Physical Sciences} \textbf{1988}, \emph{417}, 213--233\relax
\mciteBstWouldAddEndPuncttrue
\mciteSetBstMidEndSepPunct{\mcitedefaultmidpunct}
{\mcitedefaultendpunct}{\mcitedefaultseppunct}\relax
\EndOfBibitem
\bibitem[Brochard \latin{et~al.}(1984)Brochard, Jouffroy, and
  Levinson]{brochard1984phase}
Brochard,~F.; Jouffroy,~J.; Levinson,~P. Phase diagrams of mesomorphic
  mixtures. \emph{Journal de Physique} \textbf{1984}, \emph{45},
  1125--1136\relax
\mciteBstWouldAddEndPuncttrue
\mciteSetBstMidEndSepPunct{\mcitedefaultmidpunct}
{\mcitedefaultendpunct}{\mcitedefaultseppunct}\relax
\EndOfBibitem
\bibitem[Chiu and Kyu(1995)Chiu, and Kyu]{chiu1995equilibrium}
Chiu,~H.-W.; Kyu,~T. Equilibrium phase behavior of nematic mixtures. \emph{The
  Journal of Chemical Physics} \textbf{1995}, \emph{103}, 7471--7481\relax
\mciteBstWouldAddEndPuncttrue
\mciteSetBstMidEndSepPunct{\mcitedefaultmidpunct}
{\mcitedefaultendpunct}{\mcitedefaultseppunct}\relax
\EndOfBibitem
\bibitem[Kim \latin{et~al.}(2007)Kim, Choi, Chien, and Kyu]{kim2007phase}
Kim,~N.; Choi,~J.; Chien,~L.-C.; Kyu,~T. Phase equilibria of a mixture of
  side-chain liquid crystalline polymer and low molecular mass liquid crystal.
  \emph{Macromolecules} \textbf{2007}, \emph{40}, 9582--9589\relax
\mciteBstWouldAddEndPuncttrue
\mciteSetBstMidEndSepPunct{\mcitedefaultmidpunct}
{\mcitedefaultendpunct}{\mcitedefaultseppunct}\relax
\EndOfBibitem
\bibitem[Carri and Muthukumar(1998)Carri, and
  Muthukumar]{carri1998configurations}
Carri,~G.~A.; Muthukumar,~M. Configurations of liquid crystalline polymers in
  nematic solvents. \emph{The Journal of Chemical Physics} \textbf{1998},
  \emph{109}, 11117--11128\relax
\mciteBstWouldAddEndPuncttrue
\mciteSetBstMidEndSepPunct{\mcitedefaultmidpunct}
{\mcitedefaultendpunct}{\mcitedefaultseppunct}\relax
\EndOfBibitem
\bibitem[Wang and Wang(2010)Wang, and Wang]{wang2010theory}
Wang,~R.; Wang,~Z.-G. Theory of side-chain liquid crystal polymers: Bulk
  behavior and chain conformation. \emph{Macromolecules} \textbf{2010},
  \emph{43}, 10096--10106\relax
\mciteBstWouldAddEndPuncttrue
\mciteSetBstMidEndSepPunct{\mcitedefaultmidpunct}
{\mcitedefaultendpunct}{\mcitedefaultseppunct}\relax
\EndOfBibitem
\bibitem[Pasini \latin{et~al.}(2000)Pasini, Chiccoli, and
  Zannoni]{pasini2000liquid}
Pasini,~P.; Chiccoli,~C.; Zannoni,~C. \emph{Advances in the Computer
  Simulations of Liquid Crystals}; Springer, 2000; pp 121--138\relax
\mciteBstWouldAddEndPuncttrue
\mciteSetBstMidEndSepPunct{\mcitedefaultmidpunct}
{\mcitedefaultendpunct}{\mcitedefaultseppunct}\relax
\EndOfBibitem
\bibitem[Wilson \latin{et~al.}(2022)Wilson, Yu, Potter, Walker, Gray, Li, and
  Boyd]{wilson2022molecular}
Wilson,~M.~R.; Yu,~G.; Potter,~T.~D.; Walker,~M.; Gray,~S.~J.; Li,~J.;
  Boyd,~N.~J. Molecular simulation approaches to the study of thermotropic and
  lyotropic liquid crystals. \emph{Crystals} \textbf{2022}, \emph{12},
  685\relax
\mciteBstWouldAddEndPuncttrue
\mciteSetBstMidEndSepPunct{\mcitedefaultmidpunct}
{\mcitedefaultendpunct}{\mcitedefaultseppunct}\relax
\EndOfBibitem
\bibitem[Wilson(1997)]{wilson1997molecular}
Wilson,~M.~R. Molecular dynamics simulations of flexible liquid crystal
  molecules using a Gay-Berne/Lennard-Jones model. \emph{The Journal of
  Chemical Physics} \textbf{1997}, \emph{107}, 8654--8663\relax
\mciteBstWouldAddEndPuncttrue
\mciteSetBstMidEndSepPunct{\mcitedefaultmidpunct}
{\mcitedefaultendpunct}{\mcitedefaultseppunct}\relax
\EndOfBibitem
\bibitem[McBride and Wilson(1999)McBride, and Wilson]{mcbride1999molecular}
McBride,~C.; Wilson,~M.~R. Molecular dynamics simulations of a flexible liquid
  crystal. \emph{Molecular Physics} \textbf{1999}, \emph{97}, 511--522\relax
\mciteBstWouldAddEndPuncttrue
\mciteSetBstMidEndSepPunct{\mcitedefaultmidpunct}
{\mcitedefaultendpunct}{\mcitedefaultseppunct}\relax
\EndOfBibitem
\bibitem[Wilson \latin{et~al.}(2003)Wilson, Ilnytskyi, and
  Stimson]{wilson2003computer}
Wilson,~M.~R.; Ilnytskyi,~J.~M.; Stimson,~L.~M. Computer simulations of a
  liquid crystalline dendrimer in liquid crystalline solvents. \emph{The
  Journal of Chemical Physics} \textbf{2003}, \emph{119}, 3509--3515\relax
\mciteBstWouldAddEndPuncttrue
\mciteSetBstMidEndSepPunct{\mcitedefaultmidpunct}
{\mcitedefaultendpunct}{\mcitedefaultseppunct}\relax
\EndOfBibitem
\bibitem[Ilnytskyi and Neher(2007)Ilnytskyi, and Neher]{ilnytskyi2007structure}
Ilnytskyi,~J.~M.; Neher,~D. Structure and internal dynamics of a side chain
  liquid crystalline polymer in various phases by molecular dynamics
  simulations: A step towards coarse graining. \emph{The Journal of Chemical
  Physics} \textbf{2007}, \emph{126}, 174905\relax
\mciteBstWouldAddEndPuncttrue
\mciteSetBstMidEndSepPunct{\mcitedefaultmidpunct}
{\mcitedefaultendpunct}{\mcitedefaultseppunct}\relax
\EndOfBibitem
\bibitem[Ilnytskyi \latin{et~al.}(2012)Ilnytskyi, Saphiannikova, Neher, and
  Allen]{ilnytskyi2012modelling}
Ilnytskyi,~J.~M.; Saphiannikova,~M.; Neher,~D.; Allen,~M.~P. Modelling
  elasticity and memory effects in liquid crystalline elastomers by molecular
  dynamics simulations. \emph{Soft Matter} \textbf{2012}, \emph{8},
  11123--11134\relax
\mciteBstWouldAddEndPuncttrue
\mciteSetBstMidEndSepPunct{\mcitedefaultmidpunct}
{\mcitedefaultendpunct}{\mcitedefaultseppunct}\relax
\EndOfBibitem
\bibitem[Gay and Berne(1981)Gay, and Berne]{gay1981modification}
Gay,~J.; Berne,~B. Modification of the overlap potential to mimic a linear
  site--site potential. \emph{The Journal of Chemical Physics} \textbf{1981},
  \emph{74}, 3316--3319\relax
\mciteBstWouldAddEndPuncttrue
\mciteSetBstMidEndSepPunct{\mcitedefaultmidpunct}
{\mcitedefaultendpunct}{\mcitedefaultseppunct}\relax
\EndOfBibitem
\bibitem[Cleaver \latin{et~al.}(1996)Cleaver, Care, Allen, and
  Neal]{cleaver1996extension}
Cleaver,~D.~J.; Care,~C.~M.; Allen,~M.~P.; Neal,~M.~P. Extension and
  generalization of the Gay-Berne potential. \emph{Physical Review E}
  \textbf{1996}, \emph{54}, 559\relax
\mciteBstWouldAddEndPuncttrue
\mciteSetBstMidEndSepPunct{\mcitedefaultmidpunct}
{\mcitedefaultendpunct}{\mcitedefaultseppunct}\relax
\EndOfBibitem
\bibitem[Allen \latin{et~al.}(1996)Allen, Warren, Wilson, Sauron, and
  Smith]{allen1996molecular}
Allen,~M.~P.; Warren,~M.~A.; Wilson,~M.~R.; Sauron,~A.; Smith,~W. Molecular
  dynamics calculation of elastic constants in Gay--Berne nematic liquid
  crystals. \emph{The Journal of Chemical Physics} \textbf{1996}, \emph{105},
  2850--2858\relax
\mciteBstWouldAddEndPuncttrue
\mciteSetBstMidEndSepPunct{\mcitedefaultmidpunct}
{\mcitedefaultendpunct}{\mcitedefaultseppunct}\relax
\EndOfBibitem
\bibitem[Ilnytskyi \latin{et~al.}(2011)Ilnytskyi, Neher, and
  Saphiannikova]{ilnytskyi2011opposite}
Ilnytskyi,~J.~M.; Neher,~D.; Saphiannikova,~M. Opposite photo-induced
  deformations in azobenzene-containing polymers with different molecular
  architecture: Molecular dynamics study. \emph{The Journal of Chemical
  Physics} \textbf{2011}, \emph{135}, 044901\relax
\mciteBstWouldAddEndPuncttrue
\mciteSetBstMidEndSepPunct{\mcitedefaultmidpunct}
{\mcitedefaultendpunct}{\mcitedefaultseppunct}\relax
\EndOfBibitem
\bibitem[Ilnytskyi and Wilson(2001)Ilnytskyi, and Wilson]{ilnytskyi2001domain}
Ilnytskyi,~J.; Wilson,~M.~R. A domain decomposition molecular dynamics program
  for the simulation of flexible molecules with an arbitrary topology of
  Lennard--Jones and/or Gay--Berne sites. \emph{Computer Physics
  Communications} \textbf{2001}, \emph{134}, 23--32\relax
\mciteBstWouldAddEndPuncttrue
\mciteSetBstMidEndSepPunct{\mcitedefaultmidpunct}
{\mcitedefaultendpunct}{\mcitedefaultseppunct}\relax
\EndOfBibitem
\bibitem[Ilnytskyi and Wilson(2002)Ilnytskyi, and Wilson]{ilnytskyi2002domain}
Ilnytskyi,~J.~M.; Wilson,~M.~R. A domain decomposition molecular dynamics
  program for the simulation of flexible molecules of spherically-symmetrical
  and nonspherical sites. II. Extension to NVT and NPT ensembles.
  \emph{Computer Physics Communications} \textbf{2002}, \emph{148},
  43--58\relax
\mciteBstWouldAddEndPuncttrue
\mciteSetBstMidEndSepPunct{\mcitedefaultmidpunct}
{\mcitedefaultendpunct}{\mcitedefaultseppunct}\relax
\EndOfBibitem
\bibitem[Chirico and Langowski(1994)Chirico, and
  Langowski]{chirico1994kinetics}
Chirico,~G.; Langowski,~J. Kinetics of DNA supercoiling studied by Brownian
  dynamics simulation. \emph{Biopolymers: Original Research on Biomolecules}
  \textbf{1994}, \emph{34}, 415--433\relax
\mciteBstWouldAddEndPuncttrue
\mciteSetBstMidEndSepPunct{\mcitedefaultmidpunct}
{\mcitedefaultendpunct}{\mcitedefaultseppunct}\relax
\EndOfBibitem
\bibitem[Brackley \latin{et~al.}(2014)Brackley, Morozov, and
  Marenduzzo]{brackley2014models}
Brackley,~C.~A.; Morozov,~A.~N.; Marenduzzo,~D. Models for twistable elastic
  polymers in Brownian dynamics, and their implementation for LAMMPS. \emph{The
  Journal of Chemical Physics} \textbf{2014}, \emph{140}, 04B603\_1\relax
\mciteBstWouldAddEndPuncttrue
\mciteSetBstMidEndSepPunct{\mcitedefaultmidpunct}
{\mcitedefaultendpunct}{\mcitedefaultseppunct}\relax
\EndOfBibitem
\bibitem[Lequieu \latin{et~al.}(2019)Lequieu, C{\'o}rdoba, Moller, and
  De~Pablo]{lequieu20191cpn}
Lequieu,~J.; C{\'o}rdoba,~A.; Moller,~J.; De~Pablo,~J.~J. 1CPN: A
  coarse-grained multi-scale model of chromatin. \emph{The Journal of Chemical
  Physics} \textbf{2019}, \emph{150}, 215102\relax
\mciteBstWouldAddEndPuncttrue
\mciteSetBstMidEndSepPunct{\mcitedefaultmidpunct}
{\mcitedefaultendpunct}{\mcitedefaultseppunct}\relax
\EndOfBibitem
\bibitem[Thompson \latin{et~al.}(2022)Thompson, Aktulga, Berger, Bolintineanu,
  Brown, Crozier, in~'t Veld, Kohlmeyer, Moore, Nguyen, Shan, Stevens,
  Tranchida, Trott, and Plimpton]{LAMMPS}
Thompson,~A.~P.; Aktulga,~H.~M.; Berger,~R.; Bolintineanu,~D.~S.; Brown,~W.~M.;
  Crozier,~P.~S.; in~'t Veld,~P.~J.; Kohlmeyer,~A.; Moore,~S.~G.;
  Nguyen,~T.~D.; Shan,~R.; Stevens,~M.~J.; Tranchida,~J.; Trott,~C.;
  Plimpton,~S.~J. {LAMMPS} - a flexible simulation tool for particle-based
  materials modeling at the atomic, meso, and continuum scales. \emph{Computer
  Physics Communications} \textbf{2022}, \emph{271}, 108171\relax
\mciteBstWouldAddEndPuncttrue
\mciteSetBstMidEndSepPunct{\mcitedefaultmidpunct}
{\mcitedefaultendpunct}{\mcitedefaultseppunct}\relax
\EndOfBibitem
\bibitem[Becerra and Hall(2022)Becerra, and Hall]{becerra2022fixext}
Becerra,~D.; Hall,~L.~M. \url{https://github.com/hall-polymers/SCLCP},
  2022\relax
\mciteBstWouldAddEndPuncttrue
\mciteSetBstMidEndSepPunct{\mcitedefaultmidpunct}
{\mcitedefaultendpunct}{\mcitedefaultseppunct}\relax
\EndOfBibitem
\bibitem[Community(2018)]{blender}
Community,~B.~O. Blender - a 3D modelling and rendering package. Blender
  Foundation: Stichting Blender Foundation, Amsterdam, 2018\relax
\mciteBstWouldAddEndPuncttrue
\mciteSetBstMidEndSepPunct{\mcitedefaultmidpunct}
{\mcitedefaultendpunct}{\mcitedefaultseppunct}\relax
\EndOfBibitem
\bibitem[Berardi \latin{et~al.}(1998)Berardi, Fava, and
  Zannoni]{berardi1998gay}
Berardi,~R.; Fava,~C.; Zannoni,~C. A Gay--Berne potential for dissimilar
  biaxial particles. \emph{Chemical Physics Letters} \textbf{1998}, \emph{297},
  8--14\relax
\mciteBstWouldAddEndPuncttrue
\mciteSetBstMidEndSepPunct{\mcitedefaultmidpunct}
{\mcitedefaultendpunct}{\mcitedefaultseppunct}\relax
\EndOfBibitem
\bibitem[Everaers and Ejtehadi(2003)Everaers, and
  Ejtehadi]{everaers2003interaction}
Everaers,~R.; Ejtehadi,~M. Interaction potentials for soft and hard ellipsoids.
  \emph{Physical Review E} \textbf{2003}, \emph{67}, 041710\relax
\mciteBstWouldAddEndPuncttrue
\mciteSetBstMidEndSepPunct{\mcitedefaultmidpunct}
{\mcitedefaultendpunct}{\mcitedefaultseppunct}\relax
\EndOfBibitem
\bibitem[Bates and Luckhurst(1999)Bates, and Luckhurst]{bates1999computer}
Bates,~M.; Luckhurst,~G. Computer simulation studies of anisotropic systems.
  XXX. The phase behavior and structure of a Gay--Berne mesogen. \emph{The
  Journal of Chemical Physics} \textbf{1999}, \emph{110}, 7087--7108\relax
\mciteBstWouldAddEndPuncttrue
\mciteSetBstMidEndSepPunct{\mcitedefaultmidpunct}
{\mcitedefaultendpunct}{\mcitedefaultseppunct}\relax
\EndOfBibitem
\bibitem[Adams \latin{et~al.}(1987)Adams, Luckhurst, and
  Phippen]{adams1987computer}
Adams,~D.; Luckhurst,~G.; Phippen,~R. Computer simulation studies of
  anisotropic systems: XVII. The Gay-Berne model nematogen. \emph{Molecular
  Physics} \textbf{1987}, \emph{61}, 1575--1580\relax
\mciteBstWouldAddEndPuncttrue
\mciteSetBstMidEndSepPunct{\mcitedefaultmidpunct}
{\mcitedefaultendpunct}{\mcitedefaultseppunct}\relax
\EndOfBibitem
\bibitem[Emsley \latin{et~al.}(1992)Emsley, Luckhurst, Palke, and
  Tildesley]{emsley1992computer}
Emsley,~J.; Luckhurst,~G.; Palke,~W.; Tildesley,~D. Computer simulation studies
  of the dependence on density of the orientational order in nematic liquid
  crystals. \emph{Liquid Crystals} \textbf{1992}, \emph{11}, 519--530\relax
\mciteBstWouldAddEndPuncttrue
\mciteSetBstMidEndSepPunct{\mcitedefaultmidpunct}
{\mcitedefaultendpunct}{\mcitedefaultseppunct}\relax
\EndOfBibitem
\bibitem[de~Miguel \latin{et~al.}(1992)de~Miguel, Rull, and
  Gubbins]{de1992dynamics}
de~Miguel,~E.; Rull,~L.~F.; Gubbins,~K.~E. Dynamics of the Gay-Berne fluid.
  \emph{Physical Review A} \textbf{1992}, \emph{45}, 3813\relax
\mciteBstWouldAddEndPuncttrue
\mciteSetBstMidEndSepPunct{\mcitedefaultmidpunct}
{\mcitedefaultendpunct}{\mcitedefaultseppunct}\relax
\EndOfBibitem
\bibitem[Moreno-Razo \latin{et~al.}(2011)Moreno-Razo, Sambriski, Koenig,
  D{\'\i}az-Herrera, Abbott, and De~Pablo]{moreno2011effects}
Moreno-Razo,~J.~A.; Sambriski,~E.; Koenig,~G.~M.; D{\'\i}az-Herrera,~E.;
  Abbott,~N.~L.; De~Pablo,~J. Effects of anchoring strength on the diffusivity
  of nanoparticles in model liquid-crystalline fluids. \emph{Soft Matter}
  \textbf{2011}, \emph{7}, 6828--6835\relax
\mciteBstWouldAddEndPuncttrue
\mciteSetBstMidEndSepPunct{\mcitedefaultmidpunct}
{\mcitedefaultendpunct}{\mcitedefaultseppunct}\relax
\EndOfBibitem
\bibitem[Luckhurst \latin{et~al.}(1990)Luckhurst, Stephens, and
  Phippen]{luckhurst1990computer}
Luckhurst,~G.; Stephens,~R.; Phippen,~R. Computer simulation studies of
  anisotropic systems. XIX. Mesophases formed by the Gay-Berne model mesogen.
  \emph{Liquid Crystals} \textbf{1990}, \emph{8}, 451--464\relax
\mciteBstWouldAddEndPuncttrue
\mciteSetBstMidEndSepPunct{\mcitedefaultmidpunct}
{\mcitedefaultendpunct}{\mcitedefaultseppunct}\relax
\EndOfBibitem
\bibitem[Berardi \latin{et~al.}(1993)Berardi, Emersont, and
  Zannoni]{berardi1993monte}
Berardi,~R.; Emersont,~A.~P.; Zannoni,~C. Monte Carlo Investigations of a
  Gay-Berne Liquid Crystal. \emph{Journal of the Chemical Society, Faraday
  Transactions} \textbf{1993}, \emph{89}, 4069--4078\relax
\mciteBstWouldAddEndPuncttrue
\mciteSetBstMidEndSepPunct{\mcitedefaultmidpunct}
{\mcitedefaultendpunct}{\mcitedefaultseppunct}\relax
\EndOfBibitem
\bibitem[Brown \latin{et~al.}(2009)Brown, Petersen, Plimpton, and
  Grest]{brown2009liquid}
Brown,~W.~M.; Petersen,~M.~K.; Plimpton,~S.~J.; Grest,~G.~S. Liquid crystal
  nanodroplets in solution. \emph{The Journal of Chemical Physics}
  \textbf{2009}, \emph{130}, 044901\relax
\mciteBstWouldAddEndPuncttrue
\mciteSetBstMidEndSepPunct{\mcitedefaultmidpunct}
{\mcitedefaultendpunct}{\mcitedefaultseppunct}\relax
\EndOfBibitem
\bibitem[Margola \latin{et~al.}(2018)Margola, Satoh, and
  Saielli]{margola2018comparison}
Margola,~T.; Satoh,~K.; Saielli,~G. Comparison of the mesomorphic behaviour of
  1: 1 and 1: 2 mixtures of charged Gay-Berne GB (4.4, 20.0, 1, 1) and
  Lennard-Jones particles. \emph{Crystals} \textbf{2018}, \emph{8}, 371\relax
\mciteBstWouldAddEndPuncttrue
\mciteSetBstMidEndSepPunct{\mcitedefaultmidpunct}
{\mcitedefaultendpunct}{\mcitedefaultseppunct}\relax
\EndOfBibitem
\bibitem[Bobbili and Milner(2020)Bobbili, and Milner]{bobbili2020simulation}
Bobbili,~S.~V.; Milner,~S.~T. Simulation study of entanglement in semiflexible
  polymer melts and solutions. \emph{Macromolecules} \textbf{2020}, \emph{53},
  3861--3872\relax
\mciteBstWouldAddEndPuncttrue
\mciteSetBstMidEndSepPunct{\mcitedefaultmidpunct}
{\mcitedefaultendpunct}{\mcitedefaultseppunct}\relax
\EndOfBibitem
\bibitem[Stukowski(2009)]{stukowski2009visualization}
Stukowski,~A. Visualization and analysis of atomistic simulation data with
  OVITO--the Open Visualization Tool. \emph{Modelling and Simulation in
  Materials Science and Engineering} \textbf{2009}, \emph{18}, 015012\relax
\mciteBstWouldAddEndPuncttrue
\mciteSetBstMidEndSepPunct{\mcitedefaultmidpunct}
{\mcitedefaultendpunct}{\mcitedefaultseppunct}\relax
\EndOfBibitem
\bibitem[Kl{\"u}ppel(1993)]{kluppel1993characterization}
Kl{\"u}ppel,~M. Characterization of nonideal networks by stress-strain
  measurements at large extensions. \emph{Journal of Applied Polymer Science}
  \textbf{1993}, \emph{48}, 1137--1150\relax
\mciteBstWouldAddEndPuncttrue
\mciteSetBstMidEndSepPunct{\mcitedefaultmidpunct}
{\mcitedefaultendpunct}{\mcitedefaultseppunct}\relax
\EndOfBibitem
\bibitem[Ren \latin{et~al.}(2008)Ren, McMullan, and Griffin]{ren2008poisson}
Ren,~W.; McMullan,~P.~J.; Griffin,~A.~C. Poisson's ratio of monodomain liquid
  crystalline elastomers. \emph{Macromolecular Chemistry and Physics}
  \textbf{2008}, \emph{209}, 1896--1899\relax
\mciteBstWouldAddEndPuncttrue
\mciteSetBstMidEndSepPunct{\mcitedefaultmidpunct}
{\mcitedefaultendpunct}{\mcitedefaultseppunct}\relax
\EndOfBibitem
\bibitem[Mottram and Newton(2014)Mottram, and Newton]{mottram2014introduction}
Mottram,~N.~J.; Newton,~C.~J. Introduction to Q-tensor theory. \emph{arXiv
  preprint arXiv:1409.3542} \textbf{2014}, \relax
\mciteBstWouldAddEndPunctfalse
\mciteSetBstMidEndSepPunct{\mcitedefaultmidpunct}
{}{\mcitedefaultseppunct}\relax
\EndOfBibitem
\bibitem[Kremer and Grest(1990)Kremer, and Grest]{kremer1990dynamics}
Kremer,~K.; Grest,~G.~S. Dynamics of entangled linear polymer melts: A
  molecular-dynamics simulation. \emph{The Journal of Chemical Physics}
  \textbf{1990}, \emph{92}, 5057--5086\relax
\mciteBstWouldAddEndPuncttrue
\mciteSetBstMidEndSepPunct{\mcitedefaultmidpunct}
{\mcitedefaultendpunct}{\mcitedefaultseppunct}\relax
\EndOfBibitem
\bibitem[Center(1987)]{OhioSupercomputerCenter1987}
Center,~O.~S. Ohio Supercomputer Center. 1987;
  \url{http://osc.edu/ark:/19495/f5s1ph73}\relax
\mciteBstWouldAddEndPuncttrue
\mciteSetBstMidEndSepPunct{\mcitedefaultmidpunct}
{\mcitedefaultendpunct}{\mcitedefaultseppunct}\relax
\EndOfBibitem
\end{mcitethebibliography}

\end{document}